\documentclass[fleqn,usenatbib]{mnras}

\usepackage{newtxtext,newtxmath}

\usepackage[T1]{fontenc}
\usepackage{ae,aecompl}

\usepackage{graphicx}	
\usepackage{amsmath}	
\usepackage{comment}
\usepackage{pdflscape}
\usepackage{CJK}

\title[Chemical homogeneity of the Orion complex]{The GALAH survey: Chemical homogeneity of the Orion complex}

\author[J. Kos et al.]{
Janez Kos,$^{1}$\thanks{E-mail: janez.kos@fmf.uni-lj.si}
Joss Bland-Hawthorn,$^{2,3}$
Sven Buder,$^{4,3}$
Thomas Nordlander,$^{4,3}$\newauthor
Lorenzo Spina,$^{5,3}$
Kevin L. Beeson,$^{1}$
Karin Lind,$^{6,7}$
Martin Asplund,$^{8}$\newauthor
Ken Freeman,$^{4,3}$
Michael R. Hayden,$^{2,3}$,
Geraint F. Lewis,$^{2}$
Sarah L. Martell,$^{9,3}$\newauthor
Sanjib Sharma,$^{2,3}$
Gayandhi De Silva,$^{10,3}$
Jeffrey D. Simpson,$^{9,3}$
Daniel B. Zucker,$^{11,12}$\newauthor
Toma\v{z} Zwitter$^{1}$
Klemen \v{C}otar,$^{1}$
Jonti Horner,$^{13}$
\begin{CJK*}{UTF8}{gbsn}
Yuan-Sen Ting (丁源森),$^{14,15,16,4}$
\end{CJK*}\newauthor
Gregor Traven,$^{17}$\\
\\
$^{1}$Faculty of Mathematics and Physics, University of Ljubljana, Jadranska 19, 1000 Ljubljana, Slovenia\\
$^{2}$Sydney Institute for Astronomy, School of Physics, A28, The University of Sydney, NSW 2006, Australia\\
$^{3}$ARC Centre for All Sky Astrophysics in 3D, Australia\\
$^{4}$Research School of Astronomy and Astrophysics, Australian National University, Canberra, ACT 2611, Australia\\
$^{5}$School of Physics and Astronomy, Monash University, VIC 3800, Australia\\
$^{6}$Department of Astronomy, Stockholm University, AlbaNova University Centre, SE-106 91 Stockholm, Sweden\\
$^{7}$Max Planck Institute for Astronomy, K\"{o}nigstuhl 17, 69117 Heidelberg, Germany\\
$^{8}$Max Planck Institute for Astrophysics, Karl-Schwarzschild-Str. 1, D-85741 Garching, Germany\\
$^{9}$School of Physics, UNSW, Sydney, NSW 2052, Australia\\
$^{10}$Australian Astronomical Optics, Macquarie University, 105 Delhi Rd, North Ryde, 211, Australia\\
$^{11}$Department of Physics and Astronomy, Macquarie University, Sydney, NSW 2109, Australia\\
$^{12}$Macquarie University Research Centre for Astronomy, Astrophysics \& Astrophotonics, Sydney, NSW 2109, Australia\\
$^{13}$Centre for Astrophysics, University of Southern Queensland, Toowoomba, Queensland 4350, Australia\\
$^{14}$Institute for Advanced Study, Princeton, NJ 08540, USA\\
$^{15}$Department of Astrophysical Sciences, Princeton University, Princeton, NJ 08540, USA\\
$^{16}$Observatories of the Carnegie Institution of Washington, 813 Santa Barbara Street, Pasadena, CA 91101, USA\\
$^{17}$Lund Observatory, Department of Astronomy and Theoretical Physics, Box 43, SE-221 00 Lund, Sweden\\
}

\date{Accepted XXX. Received YYY; in original form ZZZ}

\pubyear{2020}

\begin{document}
\label{firstpage}
\pagerange{\pageref{firstpage}--\pageref{lastpage}}
\maketitle

\begin{abstract}
Due to its proximity, the Orion star forming region is often used as a proxy to study processes related to star formation and to observe young stars in the environment they were born in. With the release of Gaia DR2, the distance measurements to the Orion complex are now good enough that the three dimensional structure of the complex can be explored. Here we test the hypothesis that, due to non-trivial structure and dynamics, and age spread in the Orion complex, the chemical enrichment of youngest stars by early core-collapse supernovae can be observed. We obtained spectra of 794 stars of the Orion complex with the HERMES spectrograph at the Anglo Australian telescope as a part of the GALAH and GALAH-related surveys. We use the spectra of $\sim300$ stars to derive precise atmospheric parameters and chemical abundances of 25 elements for 15 stellar clusters in the Orion complex. We demonstrate that the Orion complex is chemically homogeneous and that there was no self-pollution of young clusters by core-collapse supernovae from older clusters; with a precision of 0.02 dex in relative alpha-elements abundance and 0.06 dex in oxygen abundance we would have been able to detect pollution from a single supernova, given a fortunate location of the SN and favourable conditions for ISM mixing. We estimate that the supernova rate in the Orion complex was very low, possibly producing no supernova by the time the youngest stars of the observed population formed (from around 21 to 8 Myr ago).  
\end{abstract}




\section{Introduction}

The Orion complex, at a distance of around 400~pc, is the nearest and most studied star-forming region. It serves as a proxy for the study of large, highly structured star forming regions with visible hierarchy. While most of the studies of star formation are focused into the Orion nebula cluster (ONC) and Ori A and their ongoing star formation, there are remnants of recent star formation (starting 21 Myr ago \citet{kos19}) in regions to the north and west of the ONC, and possibly in front of it \citep{alves12, kounkel17, fang17}. 

Due to its proximity, the Orion complex is the only large star forming region for which extensive, high resolution spectroscopic studies can be performed; hundreds of stars can be observed in a reasonable time. This fact, together with the interesting structure of the Orion complex (hierarchy, sequential star formation, nontrivial kinematics, unexplained origin), make it a prime case to study the chemical evolution of star forming regions. In the past, it has been observed that there are chemical inhomogeneties between stars and regions of the complex. In a series of papers \citet{cunha1, cunha2, cunha3, cunha4} analysed the abundances of Li, C, N, O, Si, and Fe, in a broad range of stellar types (18 B stars and 9 F and G stars) with great care, taking non-LTE effects into account. They discovered a trend of younger regions of the complex having higher abundances of O and Si, while the abundances of C, N, and Fe are constant \citep{cunha2}. This has been attributed to younger regions being polluted by core collapse supernovae material from older regions. Most massive core collapse supernovae dominantly produce oxygen \citep{nomoto06}, so this is indeed the expected signature of self-pollution. Others, however, do not see any correlation between age and chemical abundances in the complex \citep{diaz}, or even observe the opposite trend, at least in $\mathrm{[Fe/H]}$ \citep{biazzo11a,biazzo11b}. Such inconsistency (although the differences in absolute abundances between studies are minuscule) might well be a consequence of small number statistics. In fact, we now resolve more clusters in the Orion complex than the number of stars studied in those papers \citep{chen19, zari19}. This exposes another problem: within each region of the Orion complex are clusters of different ages, so by observing only a small number of stars, any analysis of age-abundances trends is ambiguous. Clusters, as well as individual stars, in large hierarchical star forming regions can evolve differently from very early stages of cluster formation \citep{masch10}, so a large number of stars must be observed to understand the star forming complex entirely. Larger spectroscopic surveys of low-mass stars had been performed \citep[e.g.][]{maxted08, sacco08, bayo11}, but a comparative chemical analysis of Orion regions had not been conducted with these data.

Open clusters are most commonly used to demonstrate chemical homogeneity and most show a high level of homogeneity \citep{desilva06, bovy16, casa19}. However, open clusters represent only a small fraction of the clusters that have survived past 100~Myr. Arguably, these represent the most massive clusters born in the least perturbed environments. The chemical homogeneity of their parent structures -- whole star forming regions -- is not obvious. Star forming regions can be made inhomogeneous as a result of most massive core collapse supernovae during the gravitational collapse of the cloud or could be intrinsically inhomogeneous due to their size and lack of time for the turbulence to homogenise the ISM \citep{feng}. The Orion complex is perfect for such an inquiry, as it includes the $\lambda$ Ori association, which does not appear to have had direct contact with the rest of the complex in its lifetime. There is also a relatively large age spread observed in the complex, which makes the possibility of self-pollution by core collapse supernovae real.

With modern multi-object spectrographs it is possible to observe hundreds of stars with each pointing, effectively making a complete survey of Orion complex members within the limiting magnitude range of such instruments. We use the 400 fibre HERMES instrument at the 3.9~m Anglo-Australian Telescope at the Siding Spring Observatory. Some data were taken by the GALAH survey and most by a dedicated survey performed by the GALAH team members. A similar survey was also done as part of the APOGEE 2 survey \citep{cottle18,kounkel18}. While such surveys cannot achieve the quality of dedicated star-by-star observations, the sheer quantity of data and contemporary analysis techniques can give more reliable picture of the chemical state of the complex.

In this paper we consider the hypothesis that the self-pollution in the Orion complex is possible. This is supported by: (i) a relatively large spread of ages of stars (21 Myr to 6.5 Myr in observed regions), (ii) consistent ages within clusters, (iii) non-trivial dynamics of the Orion complex, which puts older clusters into the vicinity of younger clusters at the time of their birth, and (iv) prior observations of chemical inhomogeneity, although observed in a small sample of stars. Above facts are also consistent with a theory of triggered and sequential star formation in the Orion complex \citep{lee07}. It must be noted that we did not observe the youngest regions in the complex (ONC and $\sigma$ Ori region), so our findings are based on regions Ori OB1a, OB1b, the $\lambda$ Ori association, and stars around NGC~1788. We study the chemical state and history of the Orion complex. Finding a complete history of star formation in the complex is not the scope of this paper, as we lack observations of the youngest stars and stars less massive than $0.35\ \mathrm{M_\odot}$. We also trade completeness of our target selection for a more cautious target selection, most suitable for measuring abundances of chemical elements and having high membership probabilities for identified clusters. Dynamics of the complex is not addressed, mostly for the same reasons, but is admittedly of equal importance as ages and chemical composition in figuring out the relations between clusters.

Proving that the younger Orion complex stars are polluted by supernovae ejecta from older stars would be the first time the population of polluting stars is observed alongside the polluted population. On the other hand, observational proof that large, complex, structured star forming regions with measurable intra-region age spreads are chemically homogeneous would have important implications as well. This is a proposition on which some techniques in Galactic archaeology rely on. Chemical tagging is a method by which stars from long-ago dispersed structures can be related based on similar chemical abundances. This is inevitably the destiny of the Orion star forming region as well. While some more massive open clusters can survive a few billion years, most ($>90\%$) stars are dispersed much quickly. Eventually they lose all kinematic similarity to their star forming region and can only be matched to it by their unique chemical signature. Two questions must be answered before chemical tagging of disk stars is deemed feasible: Do stars from the same star forming regions really have similar enough chemical signatures? And are we able to measure chemical abundances with sufficient precision that tens of thousands of different star forming regions can be discerned from each other \citep{ting15}? Nature and technical limitations make answering these questions difficult. There are known chemically non-homogeneous star forming regions, like $\gamma$ Vel \citep{spina14} and Orion is often pictured like that in the literature. Chemical differences have also been observed in several binaries \citep{hawkins20}. On the other hand, star forming regions, even outside the solar neighbourhood, have similar ($\pm0.15$ dex) metallicities \citep{spina17}.

Our data are described in Section \ref{sec:data}. One should also read \citet{kos17,buder18}, and \citet{buder21} for a complete overview of the GALAH survey and the data reduction. Clustering algorithm, isochrone fitting, and photometric parameters and age determination are outlined in Section \ref{sec:hr}. Additional details are found in our previous paper on the ages of the Ori OB1a association \citep{kos19}. The bulk of our procedures are described in Section \ref{sec:bayes}, where atmospheric parameters and abundances are calculated. We performed an unconventional, semi-Bayesian fit of synthetic stellar templates to observed spectra. Photometric quantities are propagated into spectral fitting and the results are probability distributions for all calculated parameters. Use of such a pedantic approach is obvious when a statistical evaluation of the chemical homogeneity is made in Section \ref{sec:abundances}. Finally, we estimate the number of core collapse supernovae in the observed population in Section \ref{sec:noumberofsn} and show that the observed IMF (initial mass function)  and good chemical homogeneity agree that there were most likely no supernovae that could have polluted the youngest populations in the Orion complex. Implications of this measurement are discussed in Section \ref{sec:conclusions}.

\section{Data}
\label{sec:data}

\begin{figure}
\centering
\includegraphics[width=\columnwidth]{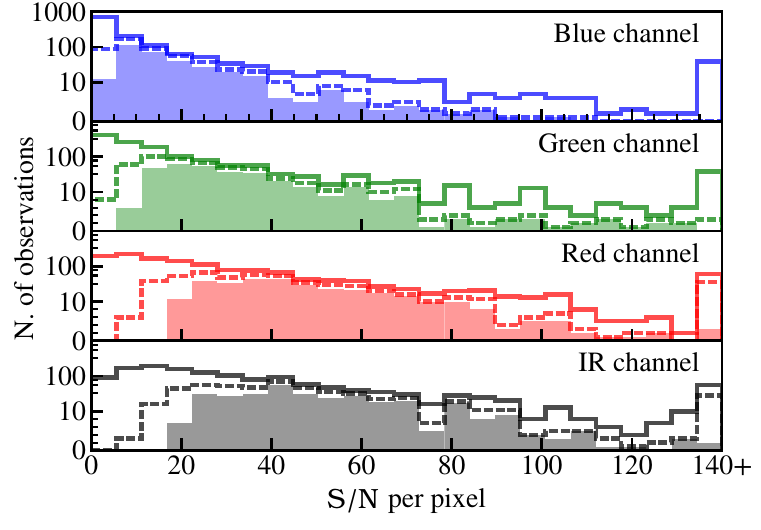}\\
\caption{Signal-to-noise ratio ($S/N$) distribution of all observed Orion stars (solid lines), spectra where the parameter pipeline converged (dashed lines) and spectra used in the final analysis (filled histograms). $S/N$ per pixel is shown. $S/N$ per resolution element is about twice as large.}
\label{fig:snr}
\end{figure}

This work relies on the observing and data reduction infrastructure of the GALAH survey. Some data were taken as part of the regular GALAH survey, but most were obtained on a separate observing proposal in order to target fainter stars and specific populations. Fields from the regular GALAH survey were observed between 2014 and 2018 and the fainter fields of the dedicated survey were observed in February 2019.

GALAH has a simple selection function, only observing stars between magnitudes $12.0<\mathrm{V_{JK}}<14.0$, where the $\mathrm{V_{JK}}$ magnitude is a $V$ magnitude calculated from 2MASS JHK photometry. A separate selection function is used for brighter targets observed during twilight, which have magnitudes $9.0<\mathrm{V_{JK}}<12.0$. Unfortunately these selection functions prevent us from observing any but the brightest A and B dwarfs in the Orion complex. While some F stars fall into the observed magnitude range, they are less likely to be Orion complex members, because observed stars are picked at random from all stars in the correct magnitude range. The GALAH selection function does not prioritise Orion members, so only a few Orion members were actually observed. To determine the abundances of a large number of elements, F, G and K type stars are more suitable than A and B stars. Hence a special survey on a separate proposal was made to observe fainter targets. Instead of using a straightforward selection function, like that for GALAH, we first found Orion complex members using the \textit{Gaia} DR2 position-proper motion-parallax space and the clustering algorithm presented in Section \ref{sec:hr}. Radial velocities were ignored at this stage and the clustering was repeated with radial velocities taken into the account once the observations were completed and all the data were reduced. Hence this initial clustering was only used to make the observing strategy as efficient as possible. Then priority was given to stars with \textit{Gaia} G magnitudes between 12.0 and 14.5 (roughly $12.25<\mathrm{V_{JK}}<14.75$). The remaining fibres were filled with Orion members up to one magnitude fainter. Orion complex members filled most of the fibre positioner's 400 fibres and any remaining fibres were positioned to capture field stars in the same magnitude range. The Ori OB1a, OB1b, $\lambda$ Ori and NGC 1788 regions were covered in the dedicated proposal. The exposure time for the fields in the separate proposal was extended by 60\% compared to GALAH fields to accommodate fainter targets. Apart from the selection function, the quality of spectra is therefore comparable in both surveys. 

Collectively, in the GALAH survey and the dedicated proposal we observed 16 fields: 11 on a separate proposal, 4 regular GALAH fields and one bright GALAH field. The bright GALAH field only includes one Orion complex member and one regular GALAH field only includes 3.  All together we observed 794 members. Most of the observed stars were not analysed fully. Final analysis of chemical homogeneity omits many stars as they are too faint for anything more than a radial velocity measurement (48\% of all observed stars). Nevertheless, these stars still help constrain the isochrone fits used for age measurements. Stars are also excluded from the final analysis if they are hotter than $T_\mathrm{eff}>7750\ \mathrm{K}$ (6\%), rotate faster than $v\,\sin\,i>40\ \mathrm{km\,s^{-1}}$ (4\%), or are double lined binary stars (1\%). Some spectra were rejected based on poor fits of spectral templates (6\%). These statistics are illustrated in Figure \ref{fig:snr}. Differences between the solid and dashed lines in Figure \ref{fig:snr} are due to hot stars (with not enough features for the pipeline to converge), binaries, fast rotators and other peculiar spectra. There are fewer stars in the final sample, as we rejected low $S/N$ spectra (with $S/N<20$ in the red arm), results with large uncertainties and moderately fast rotators ($v \sin i>40\ \mathrm{km\,s^{-1}}$). Almost 20\% of stars were observed repeatedly over an interval of years (due to the overlap between the GALAH program and the dedicated Orion observing program) or days (due to poor weather conditions during the dedicated Orion observing program).

Spectra from all observing programs cover the same wavelength range: 4718 -- 4903 {\AA} (blue channel),  5649 -- 5873 {\AA} (green channel), 6481 -- 6739 {\AA} (red channel), and 7590 -- 7890 {\AA} (infra red channel). Nominal resolving power is the same for all channels ($R=28\, 000$), but can vary between and within spectra (see Section \ref{sec:start}).

All fields/spectra were reduced with the same GALAH pipeline, regardless from which survey program they were taken. Spectra from the dedicated survey can therefore be used within the GALAH ecosystem. Any repeated observations were combined. Our analysis pipeline, however, is unique and is described in the following two sections.

\section{Clustering and ages}
\label{sec:hr}

\subsection{Clustering}

Our goal is to measure precise relative chemical abundances, which is much easier to do if measurements of individual stars can be combined to increase precision. Obviously, the measurements over a natural group of stars must be combined. The next largest structures after individual stars in the hierarchy of the complex are clusters. These do not necessarily have to be open clusters, but any reasonably large overdensities we can detect. We consider such clusters the basic building blocks of the complex; stars in each cluster are assumed to be born at the same time, in a small region. Therefore these clusters are most likely -- and indeed are assumed to be -- chemically homogeneous. Chemical abundances measured as an average over the clusters can then be measured more precisely than the abundances of individual stars.

Clusters in the Orion complex are rarely well isolated from their environment. Clustering the complex (identifying clusters within the complex) is a challenging task and is extensively explored in the literature, particularly succeeding the \textit{Gaia} DR2 \citep{kounkel18, zari19, chen19, kos19}. In general the identified clusters agree between different authors.

\begin{figure*}
    \centering
    \includegraphics[width=0.88\textwidth]{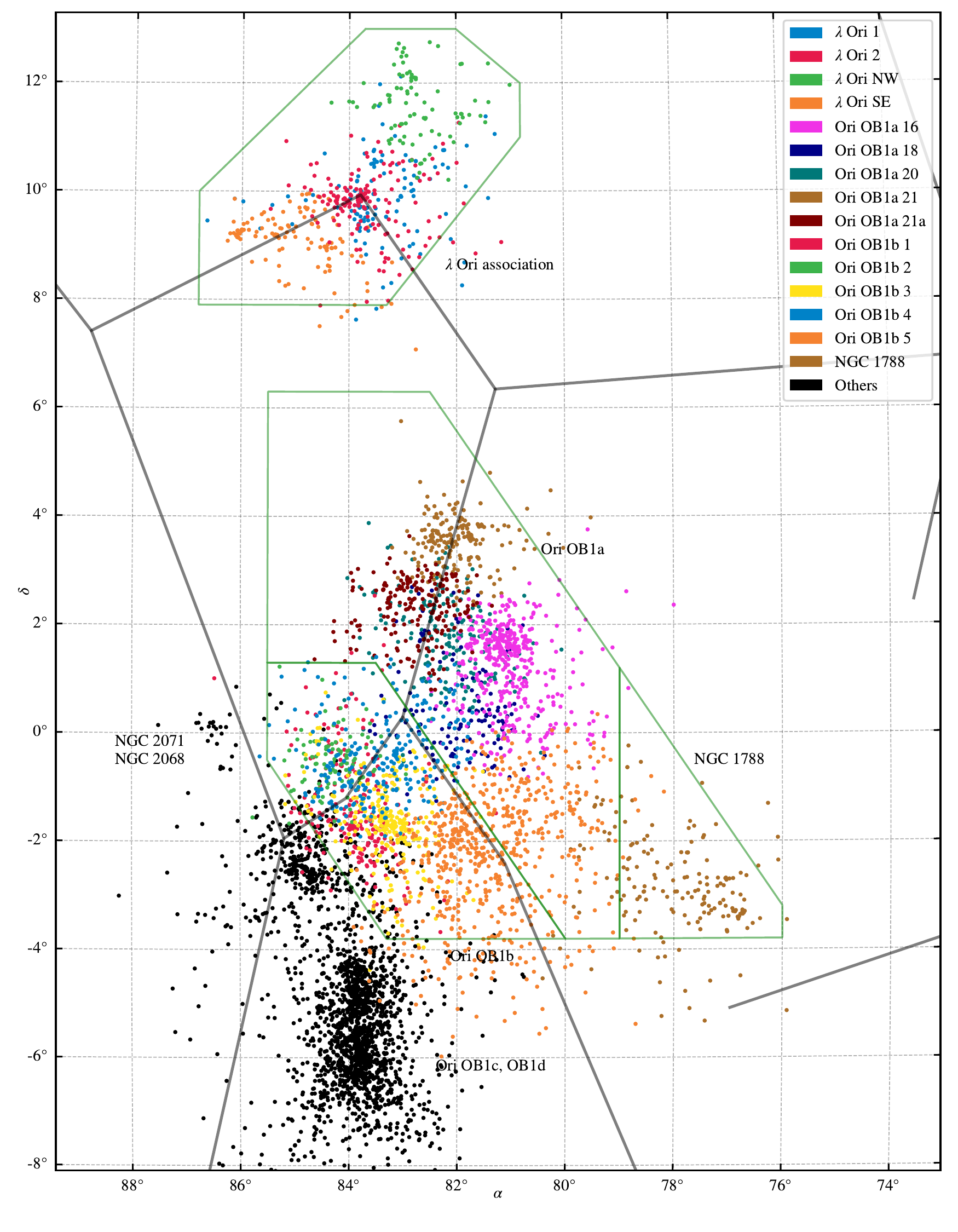}
    \caption{Orion complex with stars belonging to our clusters marked in colour. Green polygons show the region analysed in this work. 15 clusters in these regions are marked in colour. Black stars belong to other constituents of the Orion complex and are not analysed in this work. }
    \label{fig:all_clusters}
\end{figure*}

We employed a similar approach to clustering the Orion complex as in \citet{kos19}, so we only give a brief review of the method here. Parameters used in the clustering algorithm are positions, proper motions, and parallax from \textit{Gaia} DR2 \citep{gaia16,gaia18} and radial velocity, either calculated by us, or taken from \textit{Gaia} DR2 for stars not observed by us. Clusters were found using ENLINK \citep{sharma09} separately for the Ori OB1 region and the $\lambda$ Ori region. The former also included the ONC and $\sigma$ Ori cluster. In the Ori OB1 region we fixed the number of clusters to 16, as such clustering seemed plausible given the ENLINK hierarchy. 11 of them lie in our region of interest (see green polygons in Figure \ref{fig:all_clusters}). Other 5 also had to be considered, otherwise stars belonging to the $\sigma$ Ori cluster, for example, but lying close to the Ori OB1b clusters could be mis-clustered (note black points inside green polygons in Figure \ref{fig:all_clusters}). In the $\lambda$ Ori association the ENLINK clustering was more ambiguous. A small variation in parameters returned between two and 6 clusters. While two clusters are more likely, we divided the region into four clusters to check for possible chemical variations in stars close to the centre of the association as opposed to two ``tails'' stretching to the north-west and south-east.

From the ENLINK clustering we only used the centres of clusters and then found cluster members following the same approach (modified K-mean algorithm) as in \citet{kos19}: we defined a metric
\begin{multline}
\label{eq:distance}
    d=\frac{\arccos{\left( \mathbf{r} \cdot \mathbf{\overline{r}} \right)}}{1.25^\circ}+\frac{\sqrt{\left( \mu_\alpha - \overline{\mu_\alpha} \right)^2+\left(\mu_\delta - \overline{\mu_\delta} \right)^2}}{1.0\, \mathrm{mas\, yr^{-1}}}+\frac{\left | \varpi-\overline{\varpi} \right |}{0.22\, \mathrm{mas}}+\\ +\frac{\left | v_r-\overline{v_r} \right |}{15.0\, \mathrm{km\,s^{-1}}},
\end{multline}
where bars denote positions, proper motions, parallax and the radial velocity of a cluster centre. The first term describes the distance on the sky. Stars with a normalised distance $d<4.0$ from a cluster centre are made members of that cluster. If more than one cluster centre is within this distance, a star is considered to be a member of only the nearest one. If no cluster centre is within $d<4.0$ of a star it is designated a field star. Wherever no radial velocity is available, we only use the first three terms in Equation \ref{eq:distance} and scale the distance accordingly. This is described in more details in \citet{kos19}. Radial velocities used here are of similar quality as in \citet{kos19}; uncertainty of GALAH $v_r$ is around $0.25\ \mathrm{km\, s^{-1}}$, and average Gaia $v_r$ uncertainty is $4.9\ \mathrm{km\, s^{-1}}$.

After each star is assigned a cluster (or is left as a field star) we recalculate cluster centres and repeat the above process until it converges (so more than 98\% of stars do not change cluster memberships after the final iteration, approximately 5 iterations are needed). Final cluster members are illustrated in Figure \ref{fig:all_clusters}. Individual clusters in a 6D space are shown in Appendix \ref{sec:clustering_ap} and a list of members is available at CDS. The centres defining the clusters are listed in Table \ref{tab:clusters}. Clusters from \citet{kos19} are mostly the same. More radial velocity measurements are used in this paper and border regions now have some overlap with clusters in the Ori OB1b region.  

\begin{table*}
    \centering
    \setlength{\tabcolsep}{12pt}
    \begin{tabular}{lccccccc}
        \hline
        Cluster & $\alpha$ & $\delta$ & $\mu_\alpha\, \cos{\delta}$ & $\mu_\delta$ & $\varpi$ & $v_r$ & age\\[3pt]
                & $^\circ$ & $^\circ$ & $\mathrm{mas\, yr^{-1}}$ & $\mathrm{mas\, yr^{-1}}$ & $\mathrm{mas}$ & $\mathrm{km\, s^{-1}}$ & Myr\\\hline\hline
        $\lambda$ Ori 1 & 83.545 & 9.865 & 1.643 & -2.165 & 2.42 & 29.7 & 9.2$\pm$1.8\\       
        $\lambda$ Ori 2 & 83.775 & 9.844 & 0.787 & -2.097 & 2.45 & 27.6 & 6.5$\pm$1.3\\
        $\lambda$ Ori NW & 82.810 & 11.347 & 1.343 & -1.666 & 2.49 & 24.9 & 6.5$\pm$1.3\\
        $\lambda$ Ori SE & 84.577 & 9.081 & 1.440 & -2.500 & 2.50 & 27.8 & 7.0$\pm$1.4\\
        Ori OB1a 16 & 81.057 & 1.304 & 1.326 & -0.169 & 2.85 & 21.2 & 11.7$\pm$1.2\\
        Ori OB1a 18 & 81.929 & 0.317 & 0.241 & 1.174 & 2.37 & 28.3 & 12.7$\pm$1.3\\
        Ori OB1a 20 & 82.140 & 1.637 & -0.598 & 0.687 & 2.69 & 29.7 & 21.2$\pm$2.1\\
        Ori OB1a 21 & 82.052 & 3.561 & 1.432 & -0.561 & 2.86 & 20.0 & 11.0$\pm$1.1\\
        Ori OB1a 21a & 82.786 & 2.344 & 1.685 & -0.412 & 2.81 & 20.6 & 12.5$\pm$1.2\\
        NGC 1788 & 77.820 & -2.896 & 1.249 & -0.724 & 2.64 & 22.8 & 8.5$\pm$2.1\\
        Ori OB1b 1 & 83.824 & -1.594 & -1.267 & 1.048 & 2.33 & 28.8 & 17.0$\pm$3.4\\
        Ori OB1b 2 & 84.226 & -0.474 & -1.014 & -0.705 & 2.51 & 32.6 & 16.5$\pm$3.3\\
        Ori OB1b 3 & 83.192 & -1.711 & 0.051 & -0.230 & 2.36 & 30.5 & 13.0$\pm$2.6\\
        Ori OB1b 4 & 83.268 & -0.522 & 1.666 & -1.004 & 2.78 & 21.6 & 9.0$\pm$1.8\\
        Ori OB1b 5 & 81.596 & -2.029 & 1.148 & -0.910 & 2.82 & 22.5 & 11.5$\pm$2.3\\\hline
    \end{tabular}
    \caption{Parameters defining cluster centres (columns 2 --7) as used in our membership determination algorithm. We also added a column showing measured ages (not used in the membership determination algorithm).}
    \label{tab:clusters}
\end{table*}

\subsection{Isochrones fitting and ages}

We use \textit{Gaia} photometry to derive $T_\mathrm{eff}$ and $\log g$ of each star and calculate ages (see Table \ref{tab:clusters}) of clusters. We generated Padova isochrones \citep{bressan12, chen14, tang14} for the \textit{Gaia} magnitudes using the photometric system from \citet{maiz18}. Which line opacity data and models of stellar atmospheres are used to produce synthetic photometry are described in \citet{bressan12}. Age and interstellar extinction were the only free parameters. Metallicity was assumed to be $\mathrm{[M/H]}=-0.05$, consistent with the literature \citep[e.g.][]{biazzo11b}. We assumed geometric distances from \citet{bj18}. We found the best fitting isochrone by eye, same as in \citet{kos19}. Differential reddening is low in the Ori OB1a region \citep[see][]{kos19}, but significant everywhere else. Due to the lack of proper data to precisely measure the reddening of individual stars, we determined mean reddening by isochrone fitting and increased the age uncertainty for clusters in regions with higher differential reddening. One can see in Appendix \ref{sec:hr_a} that the structures of the pre-main sequence (PMS) for less massive stars, its merging into the zero age main sequence (ZAMS), and the main sequence (MS) for more massive stars are clearly visible in HR diagrams for all clusters. Hence we conclude that the differential reddening has a limited effect on measured ages. Note that only a few stars lie on the ZAMS below the PMS-ZAMS merging point. These are field stars that were not rejected by the clustering algorithm.  Once the isochrone is determined, the nearest point on the isochrone to each star gives its mass, $T_\mathrm{eff}$, $\log g$, etc. Distance to the isochrone is calculated as a minimal distance from the isochrone in a 3D magnitude space $(M_\mathrm{G}, G_\mathrm{BP}, G_\mathrm{RP})$. We are looking for a point on the isochrone at mass $m$, where the distance between the star and the isochrone is minimal: 
\begin{multline}
m=\min\\\left[\sqrt{\left(M_\textrm{G}-M_\mathrm{G}(m)\right)^2+(G_\mathrm{BP}- G_\mathrm{BP}(m))^2+(G_\mathrm{RP}-G_\mathrm{RP}(m))^2}\right]
\end{multline}
where $M_\textrm{G}$ etc. are magnitudes of stars and $M_\mathrm{G}(m)$ etc. are magnitudes on the isochrone, given as a function of mass. All other parameters in the Padova isochrones are given as a function of mass. Moreover, given the uncertainties of \textit{Gaia} magnitudes, the probability density functions (PDF) for each parameter can be acquired. Age is used later in this paper to estimate the number of supernova explosions in the observed population (Section \ref{sec:noumberofsn}). Temperature and gravity are needed to correctly marginalise measured stellar parameters over $T_\mathrm{eff}$ and $\log g$. Stars that are much closer to the binary sequence than the fitted isochrone are considered binaries. We choose to weight the distances to the binary sequence and the fitted isochrone with a factor of 0.3 in favour of the fitted isochrone. This way the stars close to the middle point are treated as single stars.

\begin{figure}
    \centering
    \includegraphics{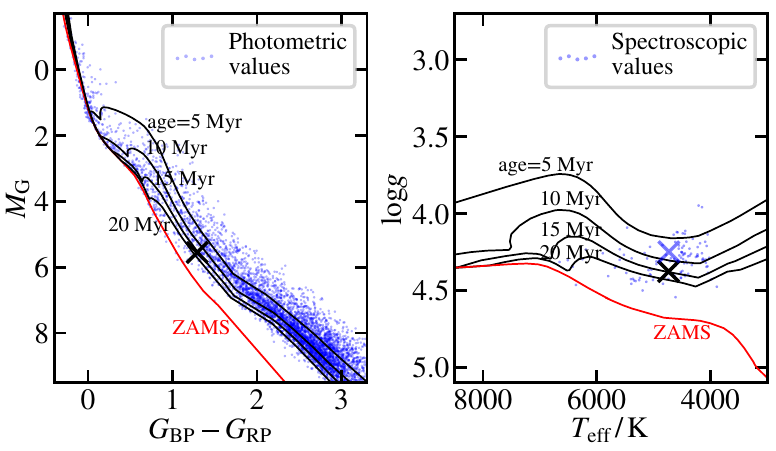}
    \caption{Left: HR diagram of all our members of the Orion complex. Right: A Kiel diagram of stars with spectroscopic $T_\mathrm{eff}$ and $\log g$. The zero-age-main-sequence (ZAMS) and isochrones for ages of 5, 10, 15, and 20 Myr are plotted, with other parameters being representative of the Orion complex. One star is marked with {\boldmath$\times$\unboldmath} in both panels to illustrate the discrepancy in spectroscopic $\log g$. In the right panel the black {\boldmath$\times$\unboldmath} indicates $T_\mathrm{eff}$ and $\log g$ calculated photometrically and the blue {\boldmath$\times$\unboldmath} indicates $T_\mathrm{eff}$ and $\log g$ measured from spectra alone.}
    \label{fig:hrkiel}
\end{figure}

$\log g$ is measured from the HR diagram much more accurately than one could from the spectra. The precision of $\log g$ in GALAH spectra is extensively discussed in \citet{buder18} and is, depending on the method used, typically worse than 0.1 dex. Temperature can be measured much more precisely in this regard. Therefore a small variation in temperature does not change the gravity measurement much (although both correlate, as seen in Figure \ref{fig:corner}). Age dependence is the exact opposite, so measuring ages well is critical for gravity estimation from the HR diagram. This is illustrated in Figure \ref{fig:hrkiel}: a difference of 0.13 dex in $log g$ is equivalent to $\sim7$ Myr (or 50\%) difference in age for a star on the 15~Myr isochrone. Given our age estimates and typical photometric uncertainties, a typical photometric $\log g$ uncertainty is 0.05~dex and a typical photometric temperature uncertainty is 60~K.

\section{Spectroscopic parameters and abundances}
\label{sec:bayes}

\subsection{Bayesian fitting schema}

The following subsection gives a general description of our approach to fitting parameters and abundances for our spectral data. Some steps are then described in more detail in Sections \ref{sec:start} to \ref{sec:end}.

\subsubsection{General description}
\label{sec:general}

To fit spectroscopic parameters and abundances we wanted to include the photometric information ($T_\mathrm{eff}$ and $\log g$) into the fitting schema. In the most basic implementation, one could leave photometric $T_\mathrm{eff}$ and $\log g$ fixed when fitting other spectroscopic parameters, but this approach has a few dangerous drawbacks. Photometric and spectroscopic parameters do not necessary represent the same quantities in practice; photometric and spectroscopic $T_\mathrm{eff}$, for example, might not measure the same temperature \citep{pins04}. Even if the definition of $T_\mathrm{eff}$ is defined consistently, different line opacity data and models of stellar atmospheres can be used for the calculation of the synthetic photometry when generating the isochrones than for the spectroscopic analysis. This can lead to large systematic errors for spectroscopic parameters. But more importantly, when aiming for the most precise chemical abundances possible, one should marginalise the calculated abundances over other measured parameters. This means that a single value for $T_\mathrm{eff}$ and $\log g$ is not sufficient, but a PDF must be used in all calculations. A PDF for $T_\mathrm{eff}$ and $\log g$ is composed from the fitted isochrone and photometric and distance uncertainties.

The above reasoning led us to adopt a Bayesian fitting scheme, where we can propagate photometrically measured $T_\mathrm{eff}$ and $\log g$ throughout the spectral fitting procedure. To fit the spectra we employ the radiative transfer code from the SME software package \citep{valenti96,piskunov17} via the iSpec wrapper \citep{cuaresma14, cuaresma19} to produce synthetic spectra. MARCS atmospheric models \citep{gustafsson08} and Gaia-ESO linelist \citep{heiter15} are used for spectrum synthesis within {iSpec}. Synthetic spectra are fitted to normalised observed spectra.

Two different fits are made. First we fit the whole spectrum in all four bands covered by the HERMES spectrograph to obtain the overall metallicity ($[M/H]$), alpha-element abundance ($[\alpha/Fe]$)\footnote{Alpha elements with lines in the covered bands are Mg, Si, Ca, Ti, and O.}, projected rotational speed ($v \sin i$), and spectroscopic $T_\mathrm{eff}$ and $\log g$. $v \sin i$ is the only fitted broadening parameter. Micro- and macro-turbulence velocities $v_\mathrm{mic}$ and $v_\mathrm{mac}$ are used in the calculation of the synthetic spectra, but are estimated by iSpec from empirical relations \citep{jofre14}. Because the observed stars are young, most are rotating fast enough that rotational broadening dominates over turbulence broadening. Elemental abundances are fitted separately and each element is fitted independently. Lines and wavelength ranges used for the fitting of elemental abundances are the same as in GALAH's DR2 \citep{buder18}. See this reference for information on the atomic data for each line.

In both cases, to fit atmospheric parameters and abundances, the log-likelihood is written as
\begin{equation}
    \ln P(f|\lambda, \sigma_f, \mathbf{\Theta})= -\frac{1}{2} \sum_n \frac{(f_n(\lambda)-s_n(\mathbf{\lambda | \Theta}))^2}{\sigma(\lambda)_f^2},
\end{equation}
where $f$ and $s$ represent the observed and synthetic spectra, the former having the uncertainty $\sigma_f$. $\mathbf{\Theta}$ are parameters of the synthetic spectrum (temperature, gravity, metallicity, etc.), and $\lambda$ is the wavelength. Summation is done over $n$ pixels or wavelength bins. The posterior probability for the fitted parameters is
\begin{equation}
    P(\mathbf{\Theta}|f,\lambda,\sigma_f) \propto P(\mathbf{\Theta}) P(f|\lambda, \sigma_f, \mathbf{\Theta}).
\end{equation}
Prior $ P(\mathbf{\Theta})$ includes all the photometric information. 

When fitting the whole spectrum the prior for $T_\mathrm{eff}$ is the PDF of the photometric temperature with the mean value corrected (see discussion on differences between photometric and spectroscopic temperature in Section \ref{sec:initial}). The prior for $\log g$ is just the PDF of the photometric gravity. Due to the proximity of the Orion complex, there is no need to improve distances by taking cluster membership into account. For the remaining parameters ($[M/H]$, $[\alpha/Fe]$, and $v \sin i$) we use flat priors; the prior probability distribution is uniform between bounds of the grid given in Table \ref{tab:grid} and zero elsewhere. Note that a separate grid is prepared for each star (see Sections \ref{sec:initial} and \ref{sec:end}). 

When fitting spectral lines of individual elements, the priors for $T_\mathrm{eff}$, $\log g$, $[M/H]$, $[\alpha/Fe]$, and $v \sin i$ are PDFs of the initial fit. This is a way to propagate global parameters to fits of individual lines, but a PDF is also needed to correctly marginalise the inferred abundances over other parameters. The PDF is represented by a multivariate Gaussian. This is a simplification, but from our experience the aforementioned PDF is indeed similar to a Gaussian and there is no visible improvement when a more complicated representation of the PDF is used.

The posterior distribution is calculated by the \textsc{emcee} code \citep{foreman}. It turns out that calculating a synthetic spectrum at every step of the Markov chain Monte Carlo algorithm (MCMC) is too time consuming. Instead we produce a grid of synthetic templates and interpolate a template at each step of the MCMC from that grid. This is much faster only if the number of spectra in a grid can be much smaller than the number of required MCMC steps. Otherwise a synthetic template spectrum should be calculated with every step of MCMC. In general, for a problem like ours, one needs $\sim50$ walkers. Based on our experimentation, around 50 steps are needed for the chains to stabilise (in the so-called burn-in phase) and tens more to sample the distribution. On top of that only $\sim20\%$ of the steps are actually accepted. These are the minimum requirements to produce useful results with well-behaved spectra. So in practice one would have to calculate on the order of $10\,000$ synthetic templates to fit one spectrum with MCMC. One can achieve a significant improvement, if a representative grid can be made from fewer synthetic spectra (see Section \ref{sec:initial}).

The results of the fitting process are PDFs for all fitted parameters and abundances. We use PDFs in the rest of our analysis whenever possible. However, sometimes mean values are used, especially to make some illustrations comprehensible.

\subsubsection{Resolution equalisation}
\label{sec:start}

The observed spectra have a nominal resolving power of $28\,000$. Actual resolving power varies with wavelength, from fibre to fibre and with time as well. Variation with wavelength is the strongest, with the resolving power dropping to around $23\,000$ in some corners of the detector. It is followed by fibre-to-fibre variations, as not all fibres produce the same sized beam and are not positioned in the pseudo slit precisely enough. The latter causes some fibre bundles to be slightly out of focus in respect to other bundles. Variations with time can also occur, if the focus of the spectrograph changes throughout the night. 

To account for varying resolution, the synthetic spectra must have the same resolution profile as the observed spectra. Synthetic spectra that can be produced at a very high resolution could be degraded to whatever is the resolution profile of the observed spectrum. This approach introduces some complications. Each observed spectrum has a different resolution profile, which requires one more operation each time a synthetic spectrum is calculated. More important is that the resolution profile is not well known. Therefore the observed spectrum and a resolution-corrected synthetic spectrum might still have relatively very different resolution profiles. 

Instead we degraded all the observed spectra so they have a constant resolution profile, with $R=22\,000$. By degrading the resolution of the observed spectra, precise knowledge of the initial resolution profile becomes less important. For a resolution degradation from $R=28\,000$ to $R=22\,000$, an uncertainty of 10\% in initial resolving power is reduced to an uncertainty of 3.9\% in the final lower resolution spectrum. 10\% uncertainty is indeed plausible for our initial spectra. Only synthetic spectra with a constant resolving power of $R=22\,000$ are needed after such an operation.

\subsubsection{Spectrum normalisation}

Eventhough the reduction pipeline provides normalised spectra, the normalisation is too crude to be used in the process described here. For this purpose we produce a synthetic spectrum with photometric $T_\mathrm{eff}$ and $\log g$, $\mathrm{[M/H]}=-0.07$, $\mathrm{[\alpha/Fe]}=0$ and $v \sin i$ estimated with iSpec. The observed and synthesised spectra are divided and the result is fitted by a high order polynomial (between orders 9 and 15, depending on the spectral band and the temperature of the star) representing the continuum. Because the spectra are expected to have similar $\mathrm{[M/H]}$ and $[\alpha/Fe]$, such a method is reliable and robust and we do not have to change the continuum at any point during the following process, not even calculating a local continuum when fitting individual lines. Normalisation is also stable for small deviations from the correct $T_\mathrm{eff}$, up to 400 K for most sensitive cold stars.

\subsubsection{Initial conditions}
\label{sec:initial}

While the MCMC algorithm itself does not need precise initial conditions, it pays to estimate all the parameters as well as possible before fitting them. The main reason is that we produce a new grid for every star and want it to be as small as possible, as long as it can contain the space sampled by MCMC. Initial conditions thus define the centre of each grid.

The initial condition for $v \sin i$ is calculated from the spectra themselves by template fitting. Even without a well known $T_\mathrm{eff}$ and metallicity one can estimate the $v \sin i$ to within a couple of $\mathrm{km\, s^{-1}}$. It is calculated in a similar way to other parameters later: a grid of synthetic spectra is calculated with different $v\sin i$ assuming the photometric temperature, $[M/H]=-0.05$ and $[\alpha/Fe]=0.0$. The grid is then interpolated, and the best matching $v \sin i$ is found. 

\begin{figure}
    \centering
    \includegraphics[width=0.95\columnwidth]{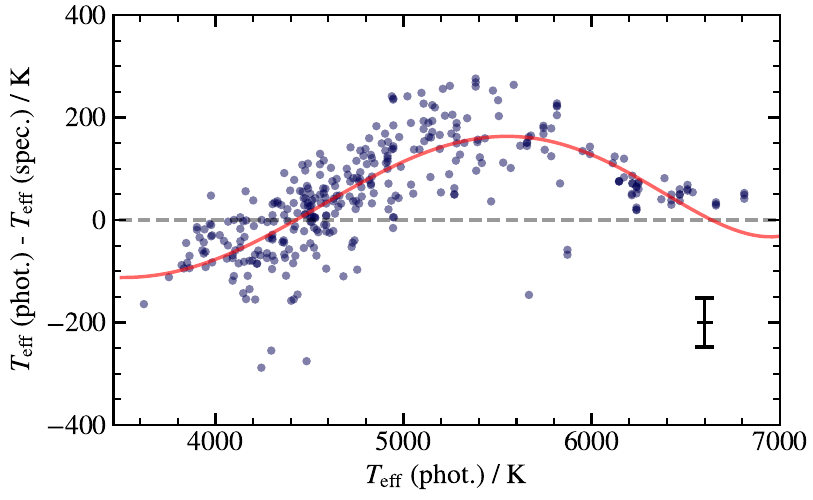}
    \caption{The difference between photometric and spectroscopic temperatures. Red line shows the relation used to construct the initial conditions for the spectroscopic temperature from its photometric counterpart. }
    \label{fig:teffdiff}
\end{figure}

The initial condition for the temperature is a slightly modified photometric temperature. We found the photometric and spectroscopic temperatures match in first order. However, there is a deviation of $\sim 160\,\mathrm{K}$ in the 4700 to 6200 K range (see Figure \ref{fig:teffdiff}). Figure \ref{fig:teffdiff} was produced by fitting the spectra with a flat prior for $T_\mathrm{eff}$. For the rest of this work we use a more restrictive prior.  The $T_\mathrm{eff}$ differences are consistent enough that we can guess in advance how different the photometric and spectroscopic temperatures will be to adjust the initial condition accordingly. Such fine tuning is not done to get a better temperature measurement or faster convergence, but to be able to make the grid as small as possible. Improvement of the initial condition by 160 K means the grid can be two or three nodes smaller in the temperature dimension, which results in a significant decrease in computing time. 

Because the Orion complex seemed to be very chemically homogeneous at first inspection, the initial conditions for metallicity and $\alpha$ abundance are $[M/H]=-0.05$ and $[\alpha/Fe]=0.0$. The initial conditions for these two parameters are not that critical, as the grid has to be relatively more extensive for them. A grid that is too small acts as a determinational prior, which we want to avoid, as metallicity and $\alpha$ abundance are the parameters we want to find. 

\subsubsection{Grid}
\label{sec:end}

As justified in Section \ref{sec:general}, it is more feasible to interpolate synthetic spectra from a small grid than producing them at every step of the MCMC algorithm. Here we explore how dense the grid must be to not introduce systematic errors into the synthetic spectra.

To evaluate how dense must the grid be, we produced Figures \ref{fig:inter_teff} -- \ref{fig:inter_vsini}. These figures show the maximum error of grid-interpolated synthetic spectra compared to synthetic spectra calculated directly by the radiative transfer code from the SME software package for the same set of parameters. Only the figures for the step sizes actually used by our analysis are shown ($\Delta T_{\mathrm{eff}}=70\, \mathrm{K}$, $\Delta [M/H]=0.05\, \mathrm{dex}$, $\Delta [\alpha/Fe]=0.05\, \mathrm{dex}$, and $\Delta(v\sin i)=2.5\, \mathrm{km\, s^{-1}}$). 

\begin{table}
    \centering
    \begin{tabular}{l@{\hskip 0.8cm}c@{\hskip 0.0cm}c @{\hskip 1.0cm}c@{\hskip 0.0cm}c}
    \hline
        \multicolumn{5}{l}{\hspace{1.6cm}Atmospheric parameters\hspace{0.8cm}Elemental abundances}\\\hline
         Parameter & \# of nodes & step size & \# of nodes & step size\\\hline\hline
         $T_\mathrm{eff}$ & $7$ & $70\ \mathrm{K}$ & $3$ & $70\ \mathrm{K}$\\
         $\log g^1$ & $3$ & $0.12\ \mathrm{dex}$ & $3$ & $0.1\ \mathrm{dex}$\\
         $v \sin i$ & $3$ & $2.5\ \mathrm{km\, s^{-1}}$ & / & /\\
         $[M/H]$ & $9$ & $0.075\ \mathrm{dex}$ & / & /\\
         $[\alpha/Fe]$ & $9$ & $0.075\ \mathrm{dex}$ & / & /\\
         $[X/Fe]$ & / & / & $30$ & $0.1\ \mathrm{dex}$\\
         \hline
         \multicolumn{5}{l}{Total \# of nodes\hspace{1cm}1701\hspace{2.9cm}270}\\
         \hline\\
         \multicolumn{5}{p{7.5cm}}{$^1$ $\log g$ dimension of the grid is omitted in practice, as marginalisation over $\log g$ had no impact on our derived PDFs (see text for explanation).}
    \end{tabular}
    \caption{Grid sizes. Atmospheric parameters and elemental abundances are fitted separately, hence two grids are needed. Parameters not used in one of the grids are marked with ``/''. A separate grid is created for each star. This table only shows the shape (dimension and resolution) of each such grid.}
    \label{tab:grid}
\end{table}

Some spectral lines seem to be very susceptible to non-linear effects and cannot be interpolated well, even with higher order splines (cubic splines were used in this work). Surprisingly the non-linear effects are limited to narrow temperature or metallicity ranges. We conclude that such phenomena are a product of SME or iSpec codes and not our interpolation (see Appendix \ref{sec:gr_interp}). These errors can be reduced by a finer grid, but not eliminated. However, a much finer grid is not feasible for our application. Such errors do not exist in the $v \sin i$ plot (Figure \ref{fig:inter_vsini}), as rotational broadening is accounted for by iSpec independently from the SME spectral synthesis code. The errors of the interpolated spectra can be neglected if they are much smaller than the uncertainty of the observed spectra (typical $S/N$ per pixel is 40, but can be as high as 100). This is true in all the cases, except for the aforementioned lines suffering from the strongest non-linear effects. However the number of such lines is small and the error is still smaller than the flux uncertainty (although not much smaller), so they have a negligible influence on the derived stellar parameters. 

As with the grid density, the grid boundaries must be as tight as possible to reduce computational time. Figure \ref{fig:corner} shows a typical PDF. The precisions of metallicity and alpha-element abundance that have otherwise non-determinant priors improve significantly when photometric priors are used. Some correlations also disappear. If the initial conditions (defining the centre of the grid) are chosen well enough, there is no need for the grid to be orders of magnitude larger than the uncertainties. Grid sizes are given in Table \ref{tab:grid}. Note that such small grids are not suitable to fit atmospheric parameters or abundances for strong outliers. They are, however, large enough to detect them. If the MCMC algorithm requires a synthetic spectrum with parameters outside the grid, a spectrum at the grid edge is returned. This effectively acts as a flat prior for all parameters. 

In the process we discovered that our results are the same if we do not marginalise the abundances PDFs over $\log g$ but instead assume photometric $\log g$ (as we always do for the initial condition). The reason is that $\log g$ can be calculated much more precisely from fitted isochrones than we ever could spectroscopically. The likelihood is essentially independent of any $\log g$ variability within the photometric $\log g$ error bars, which means that having $\log g$ as a free parameter is irrelevant. Therefore we can use grids without $\log g$, which reduces computational time significantly.

With the grid sizes discussed above, and the number of spectra in our sample, we conclude that it is more feasible to produce a small grid for each star as opposed to one giant grid spanning the parameter space of all observed stars. The grid is interpolated by cubic splines. The chosen interpolation algorithm is Scipy's \texttt{ndimage.map\_coordinates} \citep{scipy19} for its fast performance in multiple dimensions and ability to choose higher order splines as interpolation functions. 

\subsection{Evaluation of systematic effects}

\begin{figure*}
    \centering
    \includegraphics[width=0.9\textwidth]{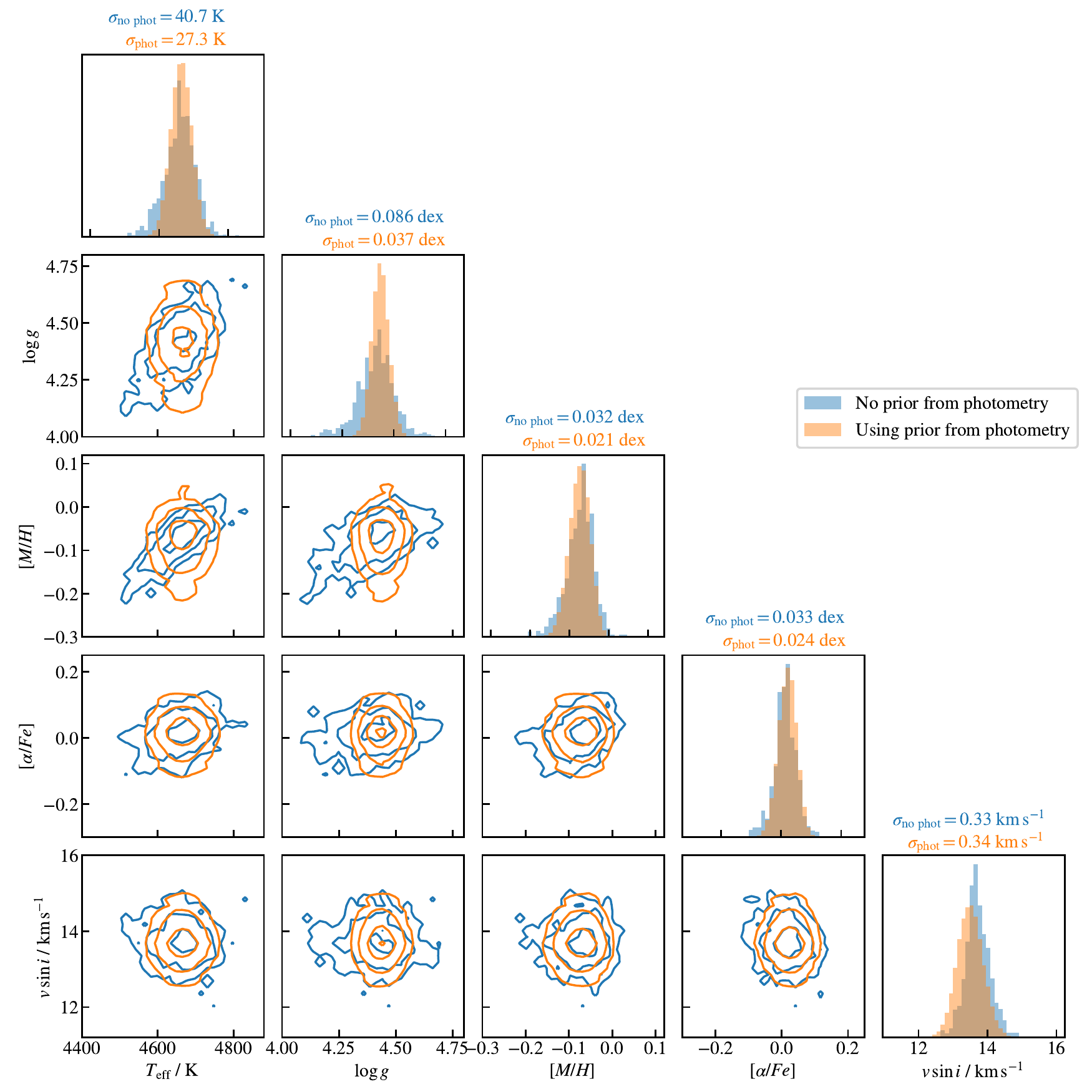}
    \caption{Corner plot showing the PDF of fitted parameters for one star without (blue) or with (orange) using priors on $T_\mathrm{eff}$ and $\log g$. Both priors are obtained from the isochrone fitting onto the HR diagram.}
    \label{fig:corner}
\end{figure*}

In Figure \ref{fig:corner} we analyse the differences between using photometric $T_\mathrm{eff}$ and $\log g$ priors in the fitting schema. The mean values for $\mathrm{[M/H]}$ and $\mathrm{[\alpha/Fe]}$ do not change much, but the uncertainty is significantly improved when priors are used. Lower uncertainty consequently has an effect on the level of measured chemical homogeneity as we compare actual PDFs and not just mean values of $\mathrm{[M/H]}$ and $\mathrm{[\alpha/Fe]}$. The uncertainty in $v \sin i$ does not improve, but the mean value does change. 

\begin{figure}
    \centering
    \includegraphics[width=\columnwidth]{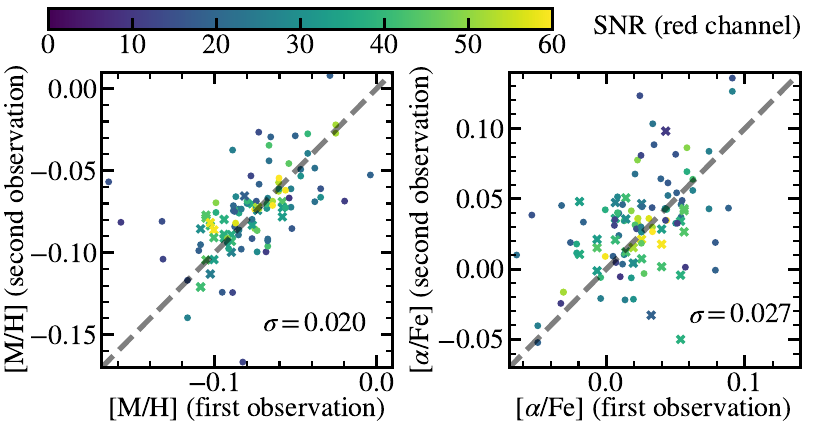}
    \caption{Analysis of repeated observations. Differences between repeats are shown for the measured metallicity (left) and alpha-element abundance (right). Circles show observations repeated with the same fibre and crosses show observations repeated with different fibres. Colour marks the lower of the two $S/N$ in the red channel. Only mean values of measured metallicities and alpha-elements abundances are shown here. Scatter around the linear relation is given in each panel.}
    \label{fig:repeats}
\end{figure}

In the fields observed in the special program we targeted members regardless of whether they were already observed in the GALAH survey. Due to poor weather we also observed some fields over several nights. Before combining observations over all epochs, we analysed individual spectra in order to estimate statistical and systematic uncertainties from repeated observations. Most observations were repeated with the same fibre (the same fields observed over several nights), but some were also done with a completely different fibre configuration (overlaps between the GALAH survey and the special program). An analysis of repeated observations is shown in Figure \ref{fig:repeats}. Metallicity and alpha-element abundance are both correlated between repeated observations, although the scatter is larger than for typical observations (see Figures \ref{fig:violin_met} and \ref{fig:violin_alpha}). The reason is that observations were repeated mostly for fields observed in poor weather conditions yielding low $S/N$. Some repeats were done for the overlap between the main GALAH program and the Orion-dedicated program. Correlation between the repeated observations is naturally better for high $S/N$ observations. We can also conclude from Figure \ref{fig:repeats} that there are no significant systematic trends related to the $S/N$ of the observation.

Uncertainties calculated by a Bayesian schema are just statistical uncertainties -- a consequence of noisy spectra, blended spectral lines, etc. Systematic uncertainties arise mostly from stars being observed with different fibres which are affected by different optical aberrations. We tried to correct for that by reducing and equalising the resolution of observed spectra, but any errors in the resolution profile are still reflected in our parameters and abundances. The scatter in metallicity and alpha-elements abundance in each cluster is larger than one would expect from statistical uncertainties alone. We attribute this to changing resolution across the CCDs, as the discrepancy between the statistical uncertainty and scatter of metallicity and alpha-elements abundance becomes lower, if only spectra with a more consistent resolution profile from the middle of the CCDs are used. This indicates that the resolution profile plays a crucial role, if very precise parameters and abundances are desired.

While the above is true for parameters measured across a wide range of wavelengths ($[M/H]$, $[\alpha/Fe]$), individual abundances suffer even more from systematic errors. Wavelength ranges where abundances are measured were carefully selected and we did not modify them from what is used in GALAH's DR2 \citep{buder18}. However a small perturbation in continuum or a nearby spectral line that might be characteristic for spectral types considered in this work can contribute some systematic uncertainty. Such contributions are very difficult to analyse and are beyond the scope of this paper. We intent to use the Orion complex and other open clusters observed in GALAH to tackle this problem in the future.

\subsection{Detrending and relative abundances}
\label{sec:abundances}

\begin{figure}
    \centering
    \includegraphics[width=0.9\columnwidth]{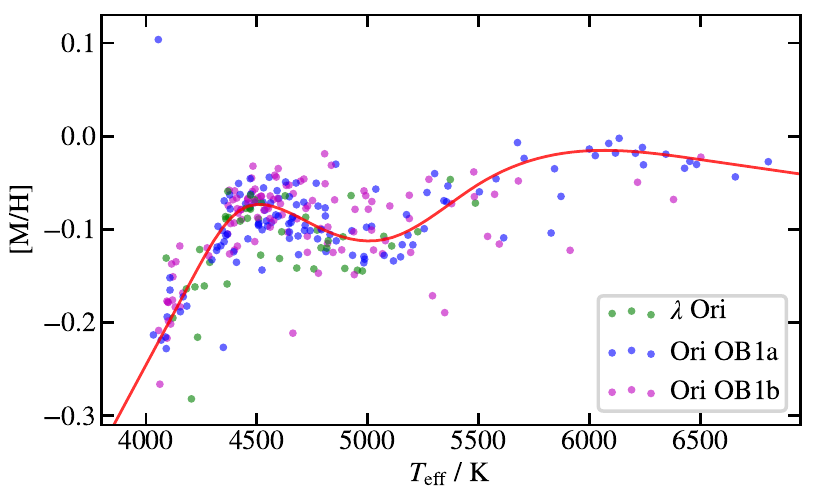}
    \caption{Measured mean metallicities for each star as a function of effective temperature. A clear trend exists and is independent of the cluster or region in the Orion complex. The solid line shows a cubic spline fit to the trend. Clusters from the Ori OB1a region are plotted in blue, from the Ori OB1b region in purple, and from the $\lambda$ Ori region in green}.
    \label{fig:detrend_mh}
\end{figure}

\begin{figure}
    \centering
    \includegraphics[width=0.9\columnwidth]{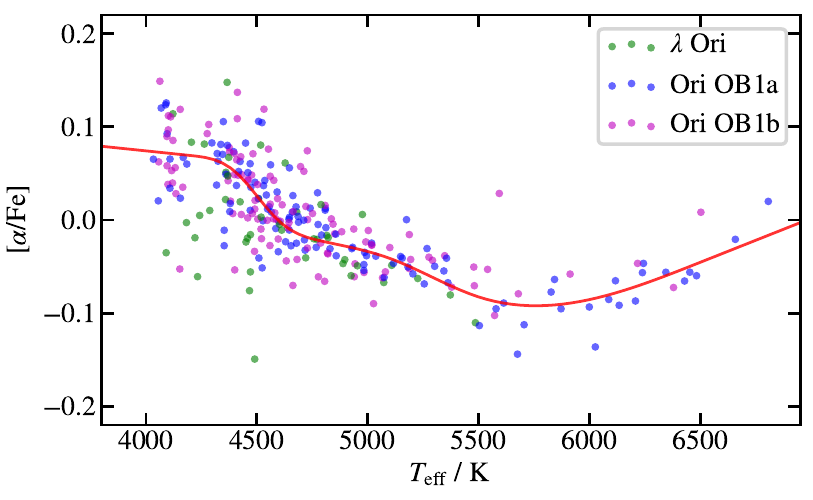}
    \caption{Same as Figure \ref{fig:detrend_mh} for alpha-element abundances.}
    \label{fig:detrend_alpha}
\end{figure}

\begin{figure*}
    \centering
    \includegraphics[width=0.95\textwidth]{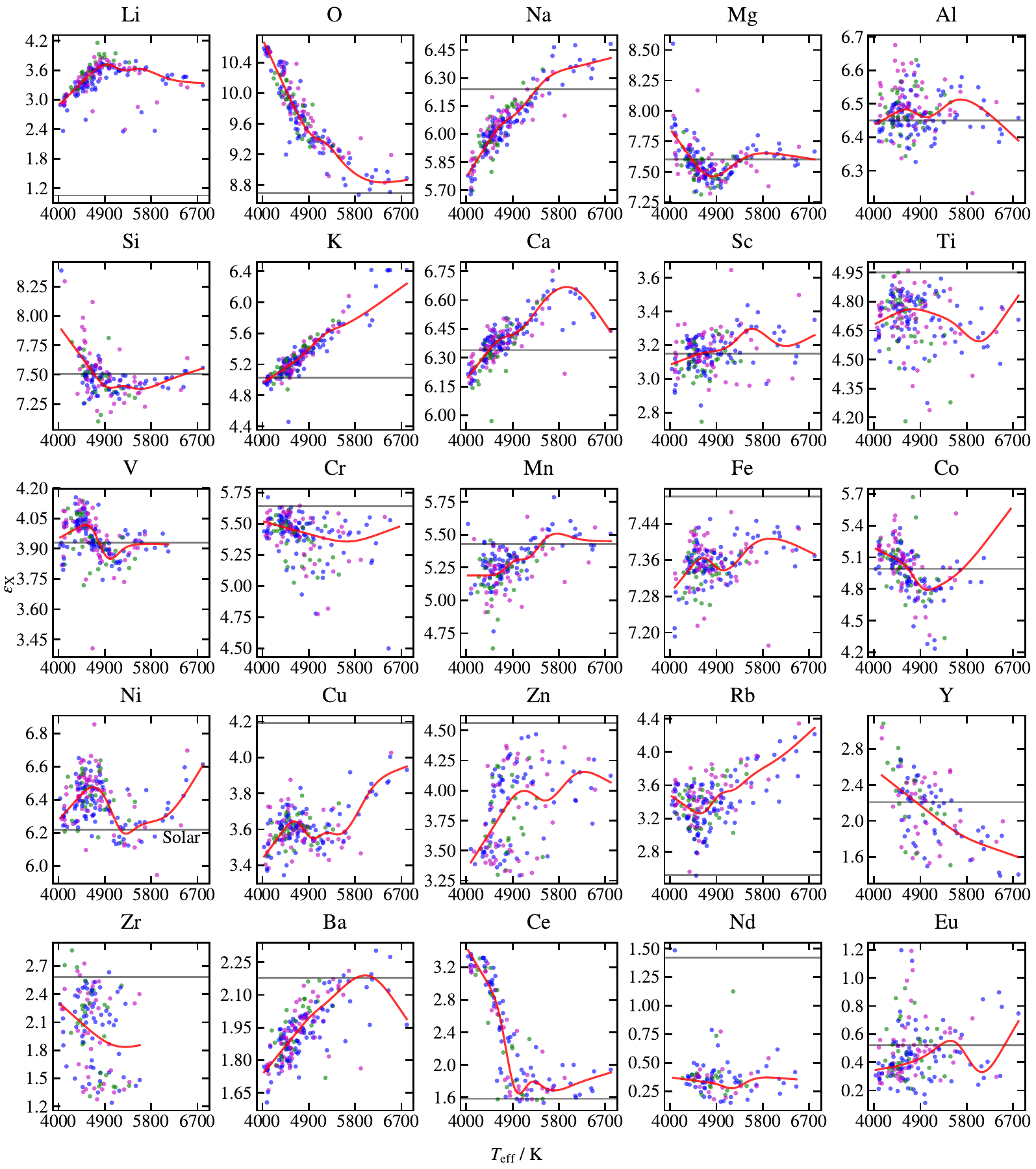}
    \caption{Measured mean abundances of 25 elements for each star as a function of effective temperature. Solid red line shows a cubic spline fit to the trend. Horizontal line shows the solar abundance from \citet{asplund09}. Clusters from the Ori OB1a region are plotted in blue, from the Ori OB1b region in purple, and from the $\lambda$ Ori region in green, same as in Figure \ref{fig:detrend_mh}.}
    \label{fig:detrend_elements}
\end{figure*}

Figures \ref{fig:detrend_mh}, \ref{fig:detrend_alpha}, and \ref{fig:detrend_elements} show the measured metallicity, alpha-elements abundance and abundances of 25 elements as a function of temperature. For metallicity and alpha elements abundance we show values normalised to the solar values\footnote{$[X/H]=\log(N_X/N_H)_\mathrm{star}-\log(N_X/N_H)_\odot$} and for elemental abundances we show absolute abundances on a $\log \epsilon$ scale\footnote{$\log \epsilon(X)=\log(N_X/N_H)+12$}.  It is evident that all parameters show trends with temperature that are the same (within our precision) for all clusters, regardless their age or location. Trends have different shapes and amplitudes for different elements. Some elements show simple trends (for example K), while other show a simple trend that plateaus off at high or low temperatures. We attribute trends to non-LTE and 3D effects and the plateau to the range of temperatures where the lines are weak and the element abundance cannot be precisely measured any more (see Ce, for example). More complicated trends are probably influenced by weak blended lines as well. In general, the following factors contribute to the trends: (i) LTE approximation. We did no correction for non-LTE or 3D effects, because we detrend all parameters anyway. Assuming the non-LTE and 3D corrections \citep{asplund05} are a smooth function of temperature, they are irrelevant after detrending. (ii) Blended lines. These are particularly important for abundances of individual elements. While only a small region around a line of interest is used to fit a model spectrum to observations, the region is not always clear of other lines. This is sometimes hard to take into the account (by changing region boundaries, for example), especially if stars with a wide range of temperatures are being compared. Large departures of abundances of some elements from the solar value is another tracer of blended lines or wrong $gf$ values. (iii) Insufficient model spectra. Model spectra cannot incorporate all physical processes. This reflects in biases we observe as trends. Some trends might be even more pronounced, as we are dealing with PMS stars, which might not have model spectra calculated as carefully and rigorously as main sequence stars. Chromospheric activity \citep{carter89, yana19} and strong magnetic fields \citep{basri92, krull99, lorenzo20} are known to influence PMS stars significantly. (iv) Biased photometric temperature and gravity could have an effect as well, although it must be minor, as this is the only effect we thoroughly analysed. 

Detrending removes any systematic and non-LTE trends well, but cannot improve the accuracy of absolute abundances. For absolute abundances we have to know the physical processes responsible for the trend. More precise absolute abundances can only be obtained by taking non-LTE effects into the account. In this work we neglect any non-LTE effects and resort to detrending. However, most of our stars are included in the GALAH DR3 \citep{buder21}, where considerable effort was put into non-LTE abundance determination. The drawback of GALAH DR3 is that the precision is lower than in this work, as stars are not assumed to be cluster members anywhere in the analysis process. Our work constrains relative chemical differences in the Orion complex much better than GALAH DR3 (see Table \ref{tab:abund} and Appendix \ref{sec:dr3}), but GALAH DR3 probably gives better mean absolute abundances. However, absolute abundances must not be always trusted, as most stars are PMS stars, which again are not treated any different to MS stars in the GALAH DR3 pipeline. Detrending can also artificially reduce the intra-cluster spread of chemical abundances, as some removed trends are physical and real. This is a drawback we can neglect, because it should impact all clusters equally, and should not affect differential abundances between clusters. 

Most notable differences between this work and GALAH DR3 are for elements Cu, Zn, Ba, and Nd. Cu and Ba probably have underestimated abundances in this work. We use lines and wavelength regions defined in GALAH DR2, but here we compare abundances with GALAH DR3, which shows higher abundances for these two elements. A direct comparison of our abundances with GALAH DR2 is impossible, because none of the stars from this paper was included in DR2. Nd abundance is measured from very weak lines and our uncertainties are underestimated, because they do not account for any errors in continuum determination. This is probably the source of large discrepancies between this work and GALAH DR3. Finally, Zn appears to show lower abundances in our work. This is one of the hardest elements to de-trend, as is obvious from Figure \ref{fig:detrend_elements}. This again means that the given uncertainties are underestimated.

For the purpose of relative chemical abundances we assume that none of the observed trends with $T_\mathrm{eff}$ and $\log g$ or $v \sin i$ are intrinsic. However, we are interested in trends with age or location, which could be a sign of chemical pollution. We do not observe trends of metallicity, alpha-elements abundance or the abundance of any of the 25 elements against any of the remaining measured atmospheric parameters ($T_\mathrm{eff}$ and $\log g$, $v \sin i$) other than temperature. Relative metallicity, alpha and elemental abundances are then calculated by removing the trend with temperature. A cubic spline is fitted as illustrated in Figures \ref{fig:detrend_mh}, \ref{fig:detrend_alpha}, and \ref{fig:detrend_elements}. Nodes were selected at an interval of $250\ \mathrm{K}$, but some were removed, so there were at least 15 data-points between each node. Three steps of a symmetric sigma clipping algorithm with a threshold of $2.5\ \sigma$ were done for the final fit.

\begin{figure*}
\centering
\includegraphics[width=0.85\textwidth]{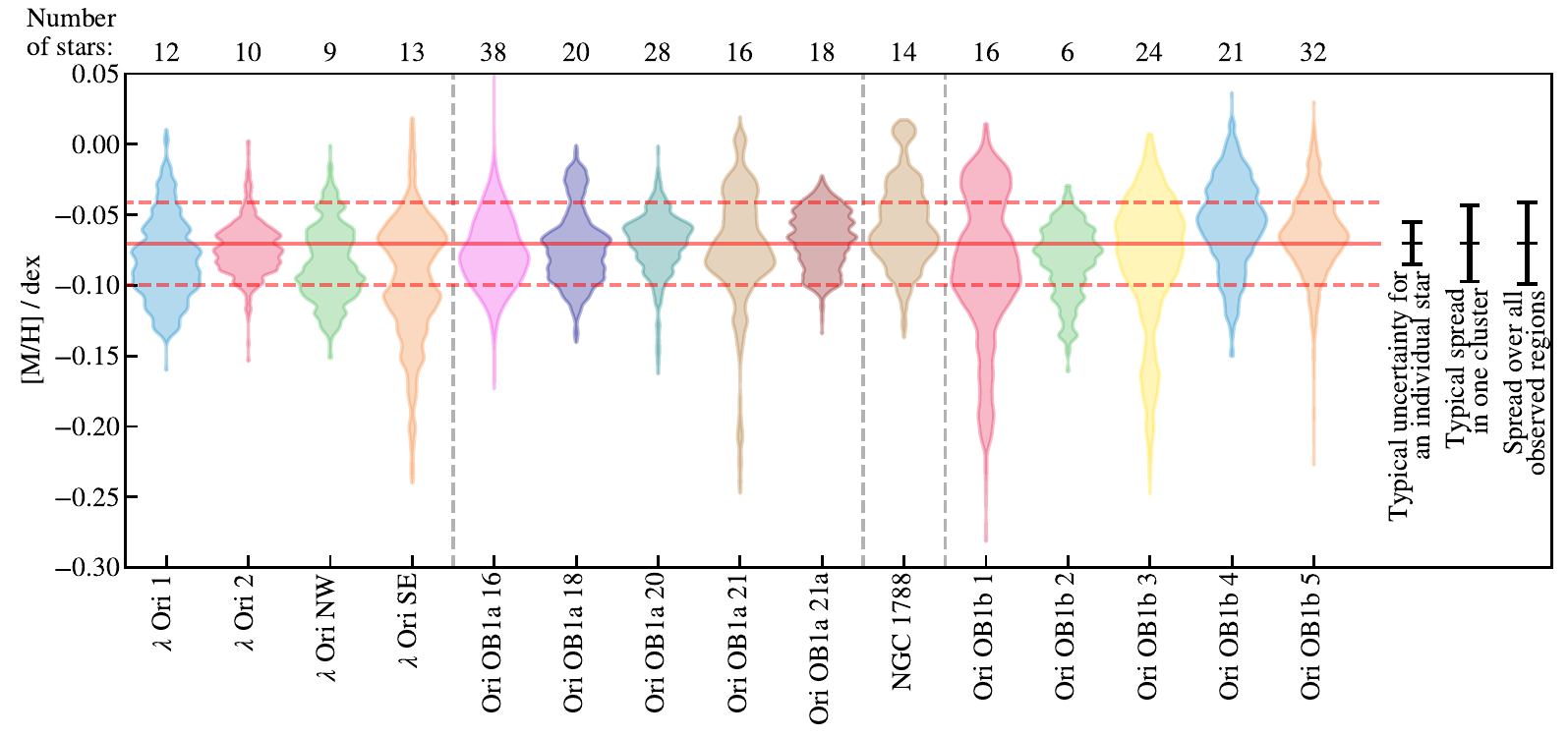}\\
\caption{The probability distribution of metallicity for all observed clusters. Each violin plot is composed of all samples of the marginalised metallicity for each analysed star. The number of stars in each cluster is given on the top. The solid horizontal line shows the mean and dashed horizontal lines show the standard deviation of the whole sample. The typical scatter of samples for individual stars is shown on the right, together with a typical scatter in one cluster and the whole region. Vertical lines divide traditional regions of the Orion complex. The colour scheme for the clusters is similar to the one used in Figure \ref{fig:all_clusters}.}
\label{fig:violin_met}
\end{figure*}

\begin{figure*}
\centering
\includegraphics[width=0.85\textwidth]{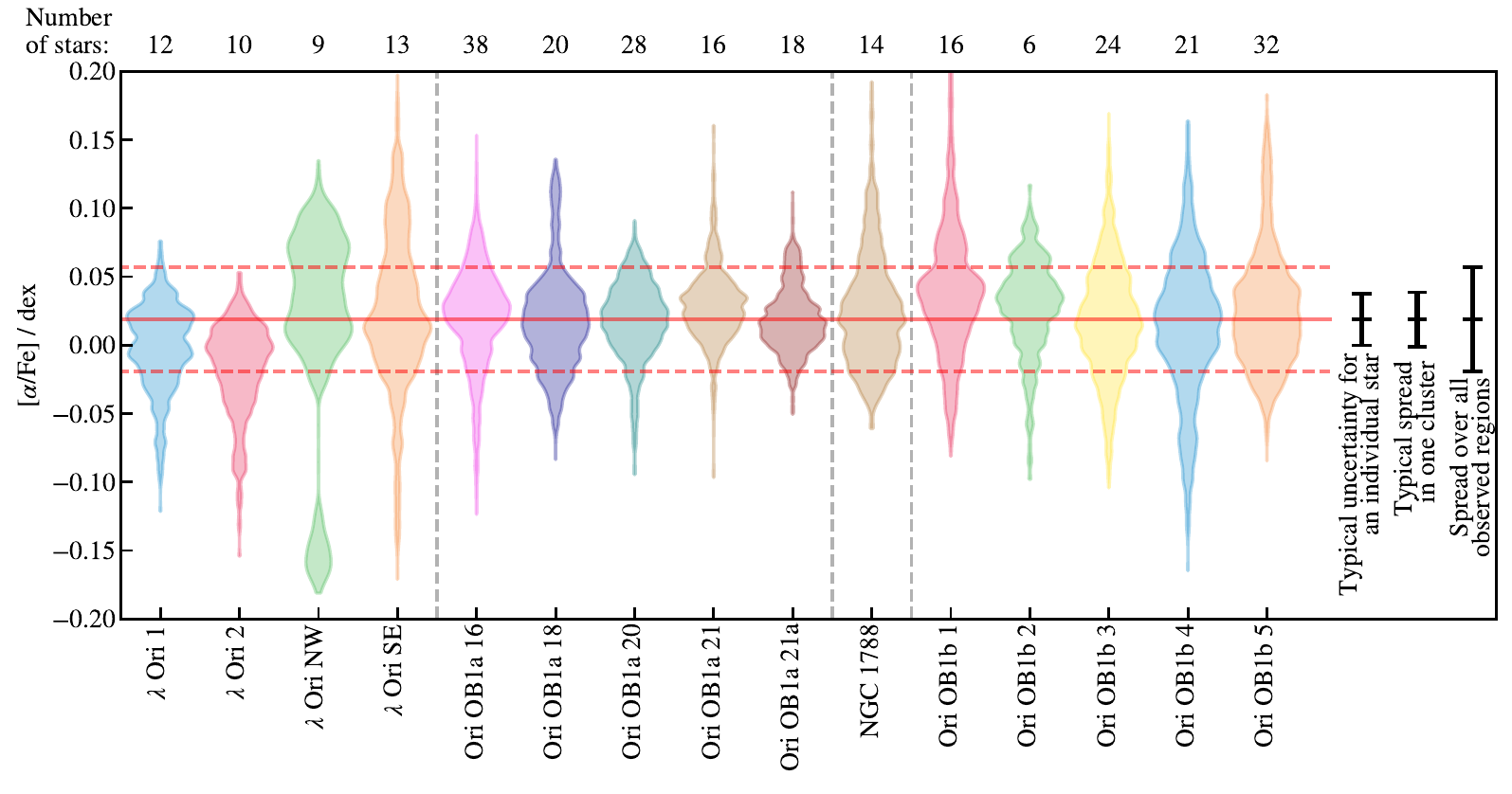}\\
\caption{Same as Figure \ref{fig:violin_met} for alpha-element abundances.}
\label{fig:violin_alpha}
\end{figure*}

A simple chi-square test shows that the observed region of the Orion complex is chemically homogeneous in metallicity, alpha-elements abundance, and all elements but lithium, whose homogeneity is not expected anyway. Li is gradually depleted early in the star's life and the abundance evolution of Li is not understood well enough to predict it at the level of homogeneity we observe here for other elements. The reduced $\chi^2$ test calculated for 15 clusters and for element $x$ is
\begin{equation}
    \chi^2(x)=\frac{1}{14}\sum_\mathrm{clusters}\frac{\left(\varepsilon_\mathrm{cluster}(x) - \overline{\varepsilon(x)} \right)^2}{\sigma_\mathrm{cluster}(x)^2+\sigma_\mathrm{int}(x)^2},
\end{equation}
where $\varepsilon_\mathrm{cluster}(x)$ is the mean abundances of element $x$ in one cluster,  $\sigma_\mathrm{cluster}(x)$ equals the measured scatter divided by the square-root of number of stars in that cluster, $\sigma_\mathrm{int}(x)$ is the intrinsic uncertainty of individual measurements. For most elements the reduced $\chi^2$ value is around 0.4, except for lithium, where it is 2.9. Detrended metallicity, alpha-elements abundance and individual elemental abundances are displayed for each cluster in Figures \ref{fig:violin_met}, \ref{fig:violin_alpha}, and \ref{fig:abund1}. 

We observe no statistically significant inhomogeneities between different clusters in Figures \ref{fig:violin_met} (metallicity) and \ref{fig:violin_alpha} (alpha-element abundance) -- all clusters have mean metallicity and alpha-element abundance within one standard deviation of the whole region. Some stars deviate by several sigmas from the mean, distorting the violin plots somewhat. These few occurrences can be explained by problems with spectrum reduction. It is also possible they are mis-identified cluster members, but the former explanation is more plausible.

There are more deviations from the mean observed in Figure \ref{fig:abund1} (elemental abundances). We again claim that outliers are are a product of reduction, as cosmic rays and telluric lines are likely to corrupt few spectral lines used in our analysis. However, more ``inhomogeneteis'' between clusters are observed in elemental abundances plots than for metallicity and alpha-element abundance. In the case of elemental abundances almost all clusters have mean abundances within one standard deviation of the whole region, with only five abundances of any cluster being up to two standard deviations from the mean. We conclude that this is not enough to claim any chemical inhomogeneity. Some clusters show a double-peaked distribution for some elements (Zr is a nice example), which can be a consequence of detrending. A physics-based detrending following non-LTE and 3D corrections might be able to solve this in the future. However, it is evident from the trends plotted in Figure \ref{fig:detrend_elements} that perhaps bimodal trends would have to be considered, which we find over-complicated for the number of observations used in this work.

\subsection{Absolute abundances}

Detrending improves the precision of our results, but not so much the accuracy of absolute abundances reported in Table \ref{tab:abund}. Therefore we report (in Table \ref{tab:abund}) abundances after they are detrended and then normalised to either the mean or the value at the solar temperature ($5770\ \mathrm{K}$). One must be careful when comparing our absolute abundances to other measurements, especially for elements that show strong trends, like O, Y, Rb, Ce, and Zr. 

\begin{figure*}
    \centering
    \includegraphics[width=\textwidth]{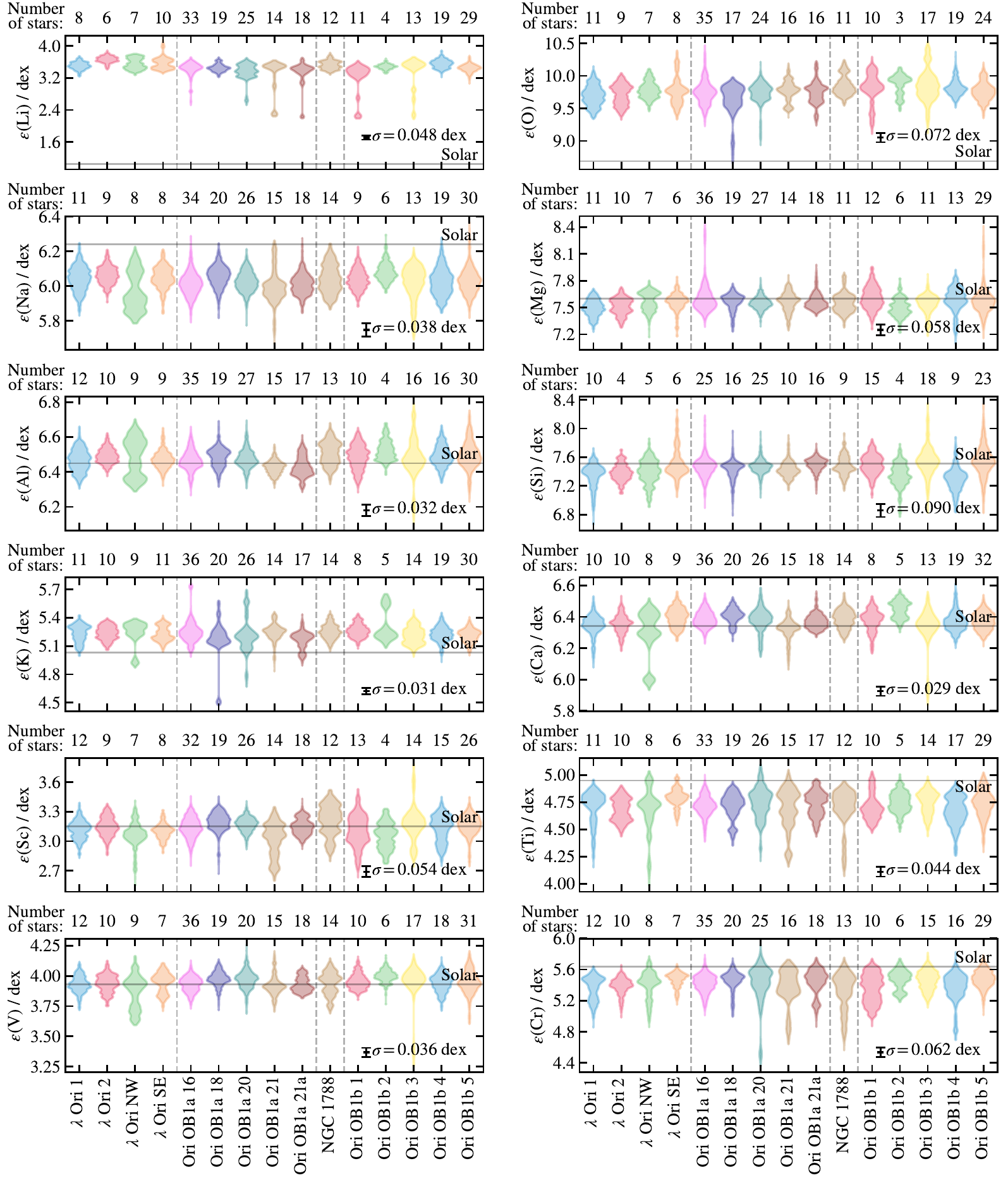}
    \caption{Probability distribution of abundances of chemical elements for all observed clusters. Each violin plot is composed of all samples of the marginalised abundance for each analysed star. Number of stars in each cluster is given on the top. Gray horizontal line shows the solar abundance. Red horizontal lines show the mean value (solid) and one standard deviation (dashed) for the abundance of each element across all clusters. Typical scatter of samples for individual stars is shown on the right. Vertical lines divide traditional regions of the Orion complex. Same colours are used for different clusters as in Figure \ref{fig:all_clusters}.}
    \label{fig:abund1}
\end{figure*}

\begin{figure*}
    \addtocounter{figure}{-1}
    \centering
    \includegraphics[width=\textwidth]{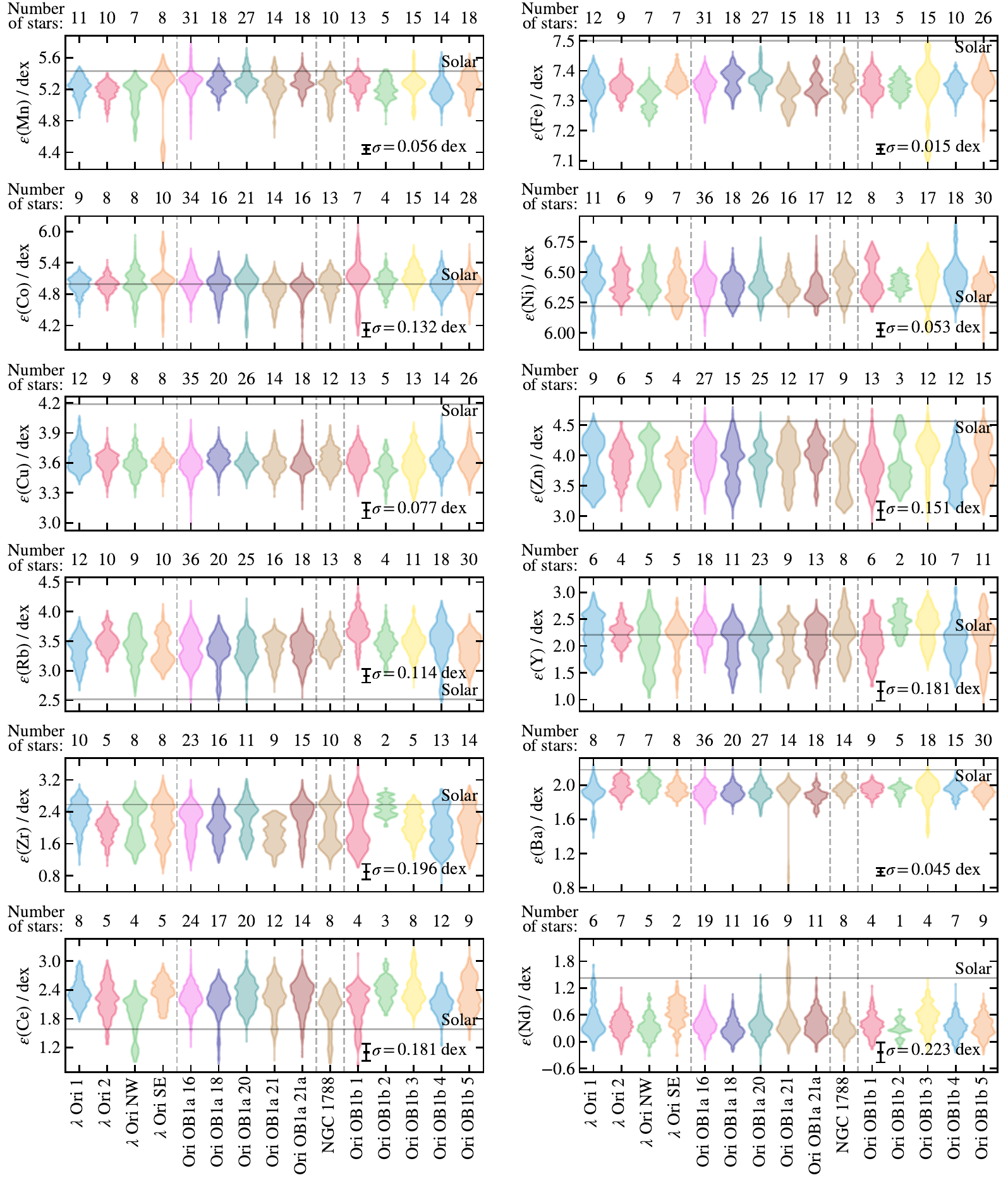}
    \caption{contd.}
\end{figure*}

\begin{figure}
    \addtocounter{figure}{-1}
    \centering
    \includegraphics[width=\columnwidth]{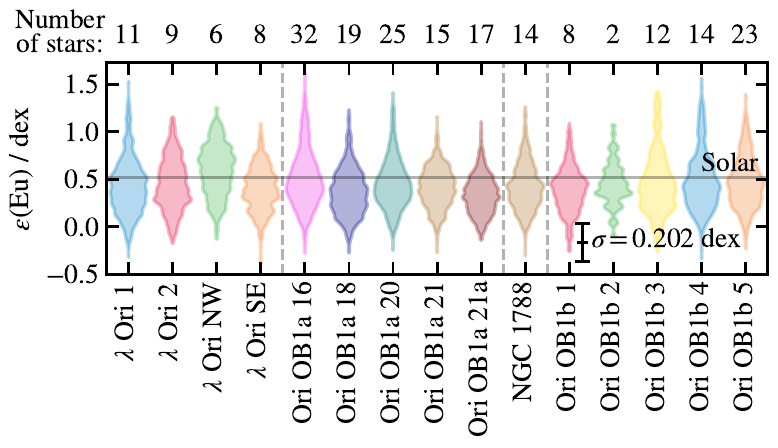}
    \caption{contd.}
\end{figure}

\begin{table*}
    \begin{tabular}{lccccp{1cm}p{0.6cm}p{1cm}p{0.6cm}cc}
    \hline
    \multicolumn{8}{c}{This work} & & \multicolumn{2}{c}{GALAH DR3}\\
    \hline
    El. & $\varepsilon(X)$ & $\varepsilon(X)$ & $\left[\frac{X}{Fe}\right]$ & $\left[\frac{X}{Fe}\right]$ & stat. uncertainty & N. of lines & notes & & $\varepsilon(X)$ & $\left[\frac{X}{Fe}\right]$\\\hline
     & mean & 5770 K & mean & 5770 K & & & & & mean & mean\\
     & normalised & normalised & normalised & normalised & & & & & normalised & normalised\\\hline\hline
Li & 3.41 & 3.55 & 2.36 & 2.50 & 0.123 & 2 & & & $3.00\pm0.10$ & $1.96\pm0.10$\\
O & 9.76 & 9.04 & 1.07 & 0.35 & 0.063 & 3 & 1 & & $9.50\pm0.14$ & $0.81\pm0.14$\\
Na & 6.02 & 6.33 & -0.21 & 0.09 & 0.025 & 3 & & & $6.14\pm0.06$ & $-0.08\pm0.06$\\
Mg & 7.57 & 7.64 & -0.02 & 0.04 & 0.040 & 3 & & & $7.48\pm0.09$ & $-0.12\pm0.08$\\
Al & 6.47 & 6.51 & 0.02 & 0.06 & 0.027 & 4 & & & $6.40\pm0.08$ & $-0.03\pm0.08$\\
Si & 7.45 & 7.37 & -0.05 & -0.13 & 0.081 & 4 & & & $7.56\pm0.06$ & $0.05\pm0.07$\\
K & 5.20 & 5.87 & 0.17 & 0.84 & 0.035 & 1 & & & $5.35\pm0.10$ & $0.27\pm0.09$\\
Ca & 6.36 & 6.68 & 0.02 & 0.34 & 0.032 & 5 & & & $6.50\pm0.08$ & $0.19\pm0.08$\\
Sc & 3.13 & 3.24 & -0.01 & 0.09 & 0.043 & 10 & & & $3.05\pm0.07$ & $-0.11\pm0.06$\\
Ti & 4.70 & 4.56 & -0.24 & -0.38 & 0.053 & 20 & & & $5.00\pm0.08$ & $0.06\pm0.07$\\
V & 3.94 & 3.91 & 0.01 & -0.01 & 0.027 & 17 & 2 & & $3.98\pm0.11$ & $0.03\pm0.10$\\
Cr & 5.41 & 5.24 & -0.22 & -0.39 & 0.054 & 9 & & & $5.64\pm0.10$ & $0.04\pm0.09$\\
Mn & 5.23 & 5.47 & -0.19 & 0.04 & 0.049 & 4 & & & $5.42\pm0.09$ & $0.03\pm0.08$\\
Fe$^*$ & 7.34 & 7.39 & -0.15 & -0.10 & 0.016 & 52 & & & $7.48\pm0.07$ & $-0.02\pm0.07$\\
Co & 4.97 & 4.99 & -0.01 & 0.00 & 0.069 & 3 & 2 & & $5.05\pm0.11$ & $-0.03\pm0.11$\\
Ni & 6.39 & 6.26 & 0.17 & 0.04 & 0.029 & 7 & & & $6.19\pm0.10$ & $-0.08\pm0.09$\\
Cu & 3.59 & 3.65 & -0.59 & -0.53 & 0.029 & 2 & & & $4.02\pm0.08$ & $-0.20\pm0.08$\\
Zn & 3.86 & 3.99 & -0.69 & -0.56 & 0.101 & 2 & & & $4.62\pm0.13$ & $0.09\pm0.13$\\
Rb & 3.38 & 3.76 & 0.86 & 1.24 & 0.097 & 1 & 1 & & $2.63\pm0.13$ & $0.018\pm0.14$\\
Y & 2.12 & 1.82 & -0.08 & -0.38 & 0.126 & 4 & 1, 3 & & $2.50\pm0.17$ & $0.38\pm0.17$\\
Zr & 2.08 & 1.70 & -0.49 & -0.87 & 0.157 & 4 & 1, 2 & & $2.90\pm0.14$ & $0.22\pm0.13$\\
Ba & 1.91 & 2.16 & -0.26 & -0.01 & 0.029 & 2 & & & $2.54\pm0.10$ & $0.39\pm0.09$\\
Ce & 2.29 & 1.67 & 0.71 & 0.09 & 0.214 & 1 & 1 & & $1.80\pm0.21$ & $0.28\pm0.21$\\
Nd & 0.37 & 0.37 & -1.04 & -1.04 & 0.099 & 5 & & & $2.42\pm0.11$ & $1.02\pm0.11$\\
Eu & 0.43 & 0.49 & -0.08 & -0.02 & 0.057 & 2 & & & $1.07\pm0.09$ & $0.54\pm0.09$\\\hline
    \end{tabular}\\
    \begin{flushleft}
    $^1$ Large range\\
$^2$ Colder stars dominate\\
$^3$ Hotter stars dominate\\
$^*$ $[Fe/H]$ instead of $[X/Fe]$ is given in 4th and 5th columns\\
\end{flushleft}
    \caption{Absolute abundances of 25 elements averaged over all stars in which each elemental abundance was measured. ``Large range'' means that the abundance measurements are spread over a large range ($>1$ dex) before detrending, so the systematic uncertainty is large. This implies that the reported absolute abundances are unreliable. Some elements have the measurements dominated by cold (note 2) or hot stars (note 3), so the mean is not calculated over the same type of stars for all elements. ``Mean normalised'' abundances were detrended so the mean remained the same after detrending. ``5770 K normalised'' abundances were detrended so the value at 5770 K (approximately solar $T_\mathrm{eff}$) remained the same. Column N. of lines gives the number of lines used to calculate elemental abundances of each element in this work.} Values in the GALAH DR3 columns are taken from \citet{buder21} and are also averaged over all stars in which each elemental abundance was measured.
    \label{tab:abund}
\end{table*}

\section{Estimating the number of supernovae}
\label{sec:noumberofsn}

To discuss the implications of a high degree of chemical homogeneity throughout the Orion complex, we want to estimate the number of SNe that exploded since the first stars in the complex were formed. One way is to integrate the IMF to calculate the expected number of massive stars that have had time to explode as core-collapse SNe.

\begin{figure}
    \centering
    \includegraphics[width=\columnwidth]{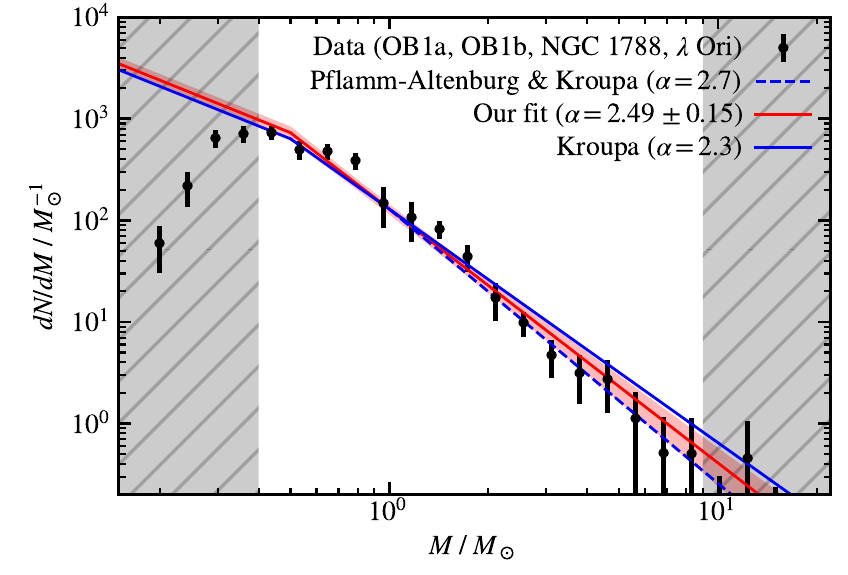}
    \caption{An estimate of the IMF of the Orion complex. Selection function is described in Section \ref{sec:noumberofsn} and is different from the selection function used for clustering. The selection is not complete in the shaded regions. A traditional Kroupa IMF shown with a solid blue line \citep{kroupa01} and an IMF obtained for massive stars by \citet{pflamm06} shown with a dashed blue line are given for reference. Our fit with $1\sigma$ uncertainty is shown in red.}
    \label{fig:imf}
\end{figure}

We estimated the masses of our members from the fitted isochrones. Because our membership selection is not complete (and is in fact quite conservative as we prioritise high membership probability to completeness and large number of targets), we also made a different selection with very relaxed criteria to be complete wherever the Gaia DR2 is. DR2 is complete between $12.0<G<17.0$ and almost complete between $7.0<G<17.0$ \citep{gaia18}. For the purpose of calculating the IMF, we selected stars in a parallax range $1.8\ \mathrm{mas}<\varpi<3.5\ \mathrm{mas},$ proper motion $\mu=\sqrt{\mu_\alpha^2+\mu_\delta^2}<3.5\ \mathrm{mas\, yr^{-1}}$, and at all magnitudes (but kept track of the completeness boundaries). Among those stars we selected all that are within $d<8.0$ of any of the 15 clusters, where $d$ is defined in Equation \ref{eq:distance}. Allowing distant stars to being members of our clusters means many stars might not have their membership determined well, but the membership of the whole complex is more complete. We also cleared the sample of any stars that can be rejected based on their position on the HR diagram, in the same manner as in the initial membership determination. Thus observed mass distribution is shown in Figure \ref{fig:imf}. We fitted a Kroupa \citep{kroupa01} IMF using data and their errorbars as seen in Figure \ref{fig:imf} to the region where our data are complete and obtained a slope $\alpha=2.49\pm0.15$. This is a bit steeper than the traditional Kroupa slope of $\alpha=2.3$ \citep{kroupa01} or $\alpha=2.27\pm0.08$ measured in the $\lambda$ Ori association \citep{nava04}, $\alpha=2.40\pm0.09$ in the 25 Ori cluster \citep{suarez19}, and $\alpha=2.4\pm$ in the ONC \citep{marchi05}, $\alpha=2.21\pm0.18$ in the Trapezium cluster \citep{muench02}, but flatter than $\alpha=2.7$ in the ONC and the Trapezium cluster \citet{pflamm06}, $\alpha=2.9\pm0.2$ in the ONC and the $\lambda$ Ori association or $\alpha=3.0\pm0.1$ in the $\sigma$ Ori cluster \citep{marchi10}.

We use data from \citet{portinari} to estimate lifetimes of stars as a function of stellar mass. A function
\begin{equation}
    \resizebox{.995\hsize}{!}{$\tau(m)=\left[3.171 \left(\frac{m}{M_\odot}\right)^{-2.178}-1.151\cdot10^{-5}\left(\frac{m}{M_\odot}\right) +0.00443\right]\,\mathrm{Gyr}$}
\end{equation}
describes the relation well for massive stars and $[M/H]=-0.07$. By using our cluster ages and extrapolating and integrating the IMF, we estimate that in the observed population there were between 0.81 and 2.28 (for $\alpha$ between $2.64$ and $2.34$) core collapse SNe in the population studied in this paper. This number drops to 0.23 -- 0.73 SNe, if we only consider time until 7 Myr ago when the last clusters formed. These estimates do not include any runaway/ejected stars into the IMF, so the actual number is a fraction higher. Assuming a steeper IMF, sometimes quoted in the literature listed above, the number of SNe drops to essentially zero. A much flatter IMF (lets say $\alpha=1.8$), which is not excluded by most massive stars ($m>5M_\odot$) and is not unprecedented in the literature \citep{marchi05,bastian10}, would produce around ten times more SNe than an IMF with $\alpha=2.49$. However, we do not expect that such an extrapolation is realistic. 

From the observation of chemical homogeneity we can deduce how many SNe would have to pollute the ISM for the younger clusters to show different chemical abundances. In the following estimate we neglect any physics of ISM mixing or cooling, as this is out of the scope of this paper. We only deal with net yields of core collapse supernovae and observed abundances. 

From Figure \ref{fig:violin_alpha} we can see that the largest mean alpha-element enhancements are 0.03 dex in the $\lambda$ Ori NW and 0.02 dex in Ori OB1b 1 clusters. However, these two clusters are unlikely to be polluted due to supernova explosions originating in older clusters. $\lambda$ Ori is too far from old clusters in Ori OB1a, so it is highly unlikely that the two groups came within a few parsecs less than 10 Myr ago (see Table \ref{tab:clusters}). Cluster Ori OB1b 1 is too old and was most likely formed before any supernova explosion took place in the Orion complex, given that the probability for a supernova in the first few million years after first star were born is low. A typical core-collapse supernova with initial mass of $25\ M_\odot$ produces $4\ M_\odot$ of alpha elements \citep{nomoto06}. Our largest clusters have masses of around $600\ M_\odot$. Assuming a star formation efficiency of 0.3 \citep{rio14}, our clusters are formed from gas clouds of around $2000\ M_\odot$. Ejecta from one $25\ M_\odot$ supernova, if completely and ideally mixed with such a cloud, would enhance its alpha-element abundance by 0.05 dex, which would be detectable in our observations. Individual abundances of Cr, Mg, and Si would increase by 0.06 dex, Ti by 0.02 dex and oxygen by 0.03 to 0.08 dex, depending which absolute abundance from Table \ref{tab:abund} is used. We measure oxygen abundances from the 777 nm oxygen triplet. These are the only oxygen lines in HERMES's spectral range and are also lines with the highest excitation potential of all fitted lines. They are known to be very sensitive to non LTE effects, chromospheric activity and atmospheric models of young stars \citep{morel04, schuler06, shen07, amarsi16}. As a result our oxygen abundances are significantly higher than in the existing literature that uses a different selection of oxygen lines \citep{cunha1, cunha2, cunha4}. Because the temperature trend (see Figure \ref{fig:detrend_elements}) is well behaved, we still use oxygen as a tracer of chemical homogeneity, but any absolute oxygen abundances given in this paper are invalid. 

The lack of any observed chemical enrichment can be explained either by no supernovae in the studied population during the star formation phase, inefficient mixing and directed flows, or large distance between the supernova and star forming regions. SNe ejecta could also be too hot and had no time to cool enough to form stars. Oxygen, which should be enriched the most, has maximum abundance ($\sim0.07$ dex above average) in clusters Ori OB1b 1, OB1b 2, and OB1b 3, which are the oldest clusters in the OB1b region. Cr, Mg, and Si are most enhanced in random clusters from any three regions with no apparent pattern. Largest enhancements are again $\sim 0.07$ dex above average, if three most enhanced clusters are compared with the rest (see Figure \ref{fig:abund1}). We can claim with high certainty that the number of SNe in the Orion complex was not high, as this would be reflected in more consistent chemical inhomogeneities.

\section{Conclusions and discussion}
\label{sec:conclusions}

In this work, we consider the question of whether the Orion star forming complex might exhibit evidence of contamination by supernovae in the elemental abundances of different clusters that formed at different times in the complex's history. To do this, we analyse data obtained using the HERMES instrument on the Anglo-Australian telescope by the GALAH survey, and additional data focusing specifically on the Orion complex obtained by the same instrument during a series of special programs. We find that the various clusters distributed across the complex are chemically homogeneous, with the younger clusters showing no evidence of pollution from supernovae amongst the older clusters.

Our conclusions are based on the Ori OB1a, OB1b, $\lambda$ Ori and NGC~1788 regions. We did not observe the ONC or $\sigma$ Ori regions due to a number of reasons. These are the regions with the largest differential reddening, which we have no way of properly accounting for. This would result in poorly fitted isochrones and consequently inaccurate photometric temperatures and gravities. We did not analyse how this would impact our spectroscopic analysis, but our approach would definitely have had to be revised to include younger stars and clusters with strong differential reddening. Stars that were included in the analysis are either main sequence stars or PMS stars very close to the main sequence. Stars $<5\ \mathrm{Myr}$ old, like the ones in ONC and $\sigma$ Ori regions would be PMS stars, lying well above the main sequence, and we had concerns that synthetic model codes would not perform well enough on them. This could make calculation of precise relative abundances hard and we would not be able to properly interpret any observed chemical differences. However, we did observe one star that is more likely to belong to the $\sigma$ Ori cluster and has well determined atmospheric parameters and abundances. It is chemically identical to other stars from this work.

Despite the Orion complex being chemically homogeneous in this study as opposed to some earlier studies, our absolute abundances of chemical elements and metallicity  agree well with the literature \citep{cunha1, cunha2, cunha3, cunha4, diaz, biazzo11a, biazzo11b}. A notable exception is oxygen, for which we were unable to calculate accurate absolute abundances. For oxygen, the measurements in the literature are done on lines with low excitation potential are thus more accurate.

The number of supernovae that have exploded to date in the Orion complex is a highly debated topic. Bubble-like structures (Barnard's loop, the Orion-Eridanus superbubble, the $\lambda$ Orion bubble) were most likely made by supernova explosions, but the present structure suggests that stellar winds played a significant role as well \citep{ochsendorf15}, leading to the formation of a rich substructure. The expansion velocity of the bubbles can be used as an indicator when they were formed -- but such estimates are inaccurate \citep{bally08} and cannot provide the exact time of the supernova explosions. \citet{brown95} estimate the age of the largest bubble -- the Orion Eridanus superbubble -- is between 1.8 and 5.3 Myr. This suggests that the bubbles are a product of recent supernovae explosions (more recent than the time of formation of the youngest stars studied in this paper). A low number of supernovae in the Orion complex is also supported by a shortfall of supernovae remnants. G203.2 - 12.3 is the only supernova remnant classified in Orion \citep[possibly observed in 483 CE]{winkler92}. Another indirect tracer is emission from the radioactive decay of $^{26}\mathrm{Al}$ in Ori OB1a \citep{voss10, schlafly15}. Models of SN and stellar wind feedbacks by \citet{voss10} suggest that $^{26}\mathrm{Al}$ was produced by a few recent supernovae. $^{60}\mathrm{Fe}$ is another radioactive tracer of SN \citep{wang20}, which has not been explored yet in the Orion complex.

More recent studies of the formation of the Orion complex are based on new Gaia distances and 3D velocities and offer a compelling picture. The observed 3D kinematics can be explained with a ``few or several'' supernovae assuming no other forces \citep{josefa20}. The most major disruption event is thought to have happened 6 Myr ago \citep{kounkel20, josefa20}, which supersedes the creation of stars studied in this work.

Our results strongly suggest that there were no or at least very few supernovae explosions in the early stages of the Orion complex formation. Young supernova remnants, also in the form of gas bubbles, can be explained by recent supernovae in the past few million years. Such supernovae are younger than observed stars and could not have chemically polluted the ISM from which these stars were born.

A convincing way of proving the chemical homogeneity of clusters is a direct comparison of spectra \citep{bovy16}. The method avoids calculating atmospheric parameters and deriving exact chemical abundances. In our case the parameter space is too large and we would struggle to find spectra with similar atmospheric parameters in order to compare lines of interesting elements. One of the reasons this is extremely difficult in young stellar associations (as compared to old open clusters) is that stellar rotation can have a large range of values, which effectively adds another dimension of atmospheric parameters.  Another drawback is that clusters of different ages would have to be compared, complicating the case for direct spectral comparison even further.

\section*{Acknowledgements}

Authors are thankful to the referee for a plenty of useful comments. This work is based on data acquired through the Australian Astronomical Observatory, under programmes: A/2019A/01 (Hierarchical star formation in Ori OB1), A/2014A/25, A/2015A/19, A2017A/18 (The GALAH survey); A/2015A/03, A/2015B/19, A/2016A/22, A/2016B/12, A/2017A/14 (The K2-HERMES K2-follow-up program); A/2016B/10 (The HERMES-TESS program); A/2015B/01 (Accurate physical parameters of Kepler K2 planet search targets); S/2015A/012 (Planets in clusters with K2). We acknowledge the traditional owners of the land on which the AAT stands, the Gamilaraay people, and pay our respects to elders past and present. JDS acknowledges the support of the Australian Research Council through Discovery Project grant DP180101791. SLM acknowledges support from the UNSW Scientia Fellowship program, and from the Australian Research Council through Discovery Project grant DP180101791. YST is supported  by the NASA Hubble Fellowship grant HST-HF2-51425.001 awarded by the Space Telescope Science Institute. This work has made use of data from the European Space Agency (ESA) mission \textit{Gaia} (\url{https://www.cosmos.esa.int/gaia}), processed by the \textit{Gaia} Data Processing and Analysis Consortium (DPAC, \url{https://www.cosmos.esa.int/web/gaia/dpac/consortium}). Funding for the DPAC has been provided by national institutions, in particular the institutions participating in the \textit{Gaia} Multilateral Agreement.

\section*{Data availability}

The data underlying this article are available from CDS, at \url{https://vizier.u-strasbg.fr/viz-bin/VizieR} and in online supplementary material.




\bibliographystyle{mnras}
\bibliography{bib}



\clearpage
\appendix

\section{Clustering}
\label{sec:clustering_ap}

Figure \ref{fig:clusters_a} shows each cluster in 6D space (position on the sky, proper motions and parallax).

\begin{figure*}
\centering
\includegraphics[width=0.64\textwidth]{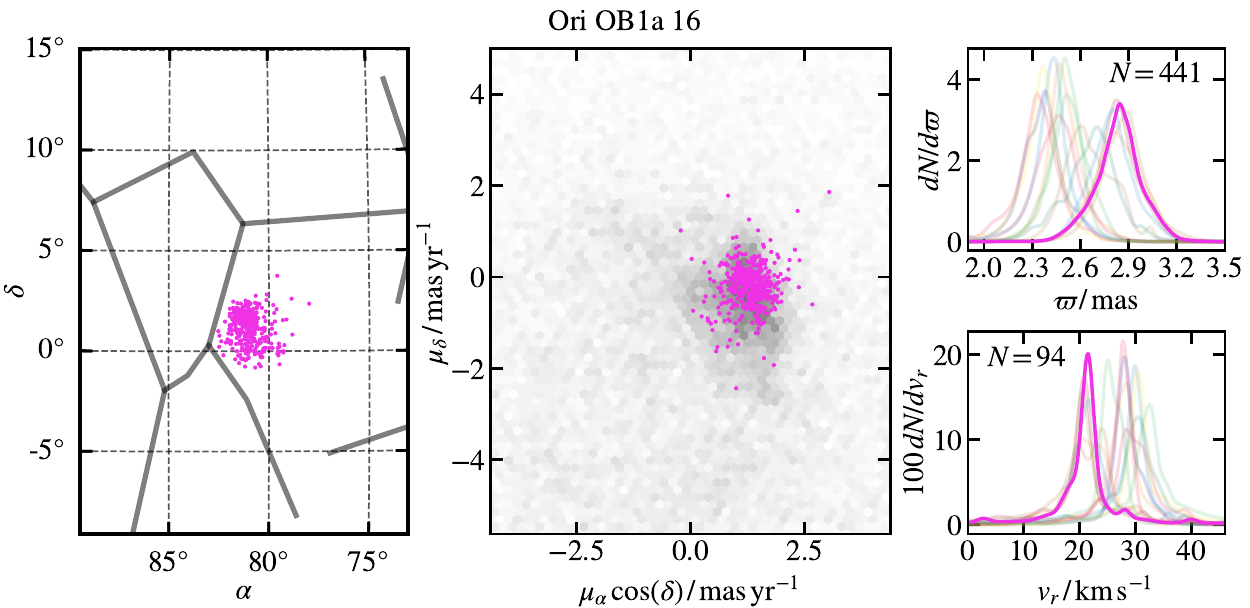}\\[0.3cm]
\includegraphics[width=0.64\textwidth]{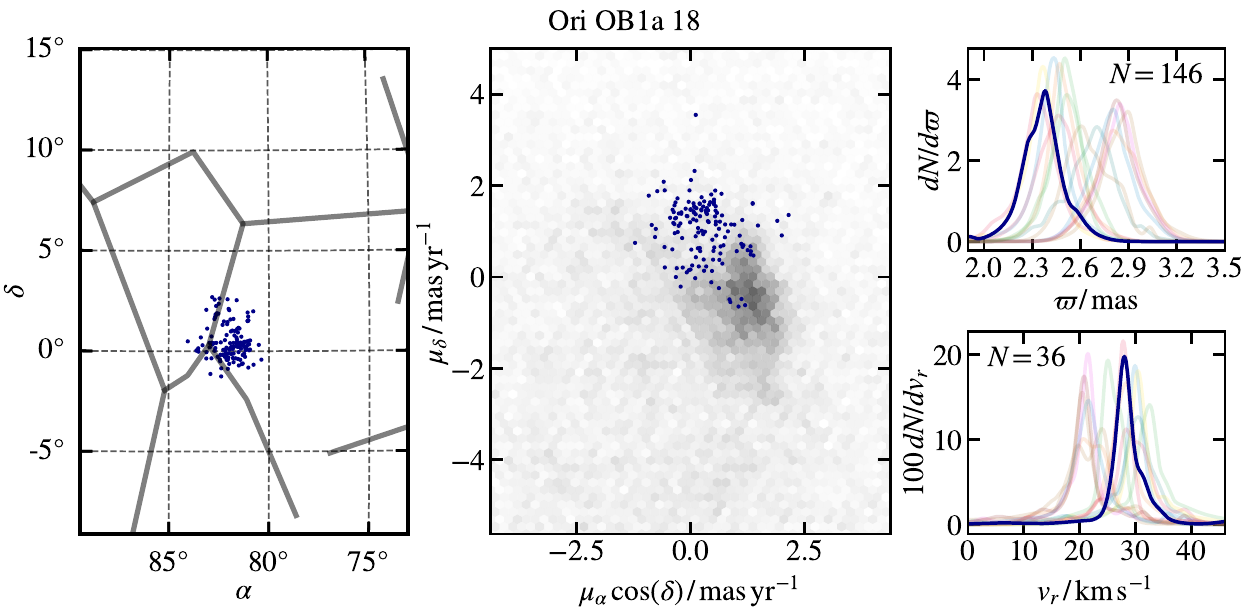}\\[0.3cm]
\includegraphics[width=0.64\textwidth]{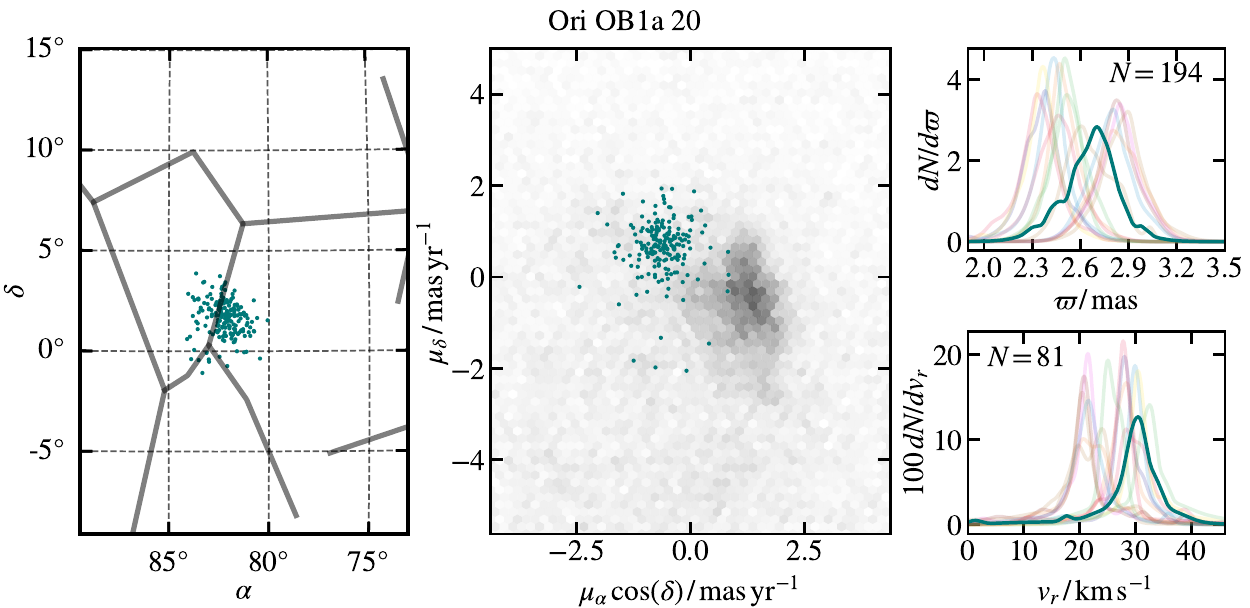}\\[0.3cm]
\includegraphics[width=0.64\textwidth]{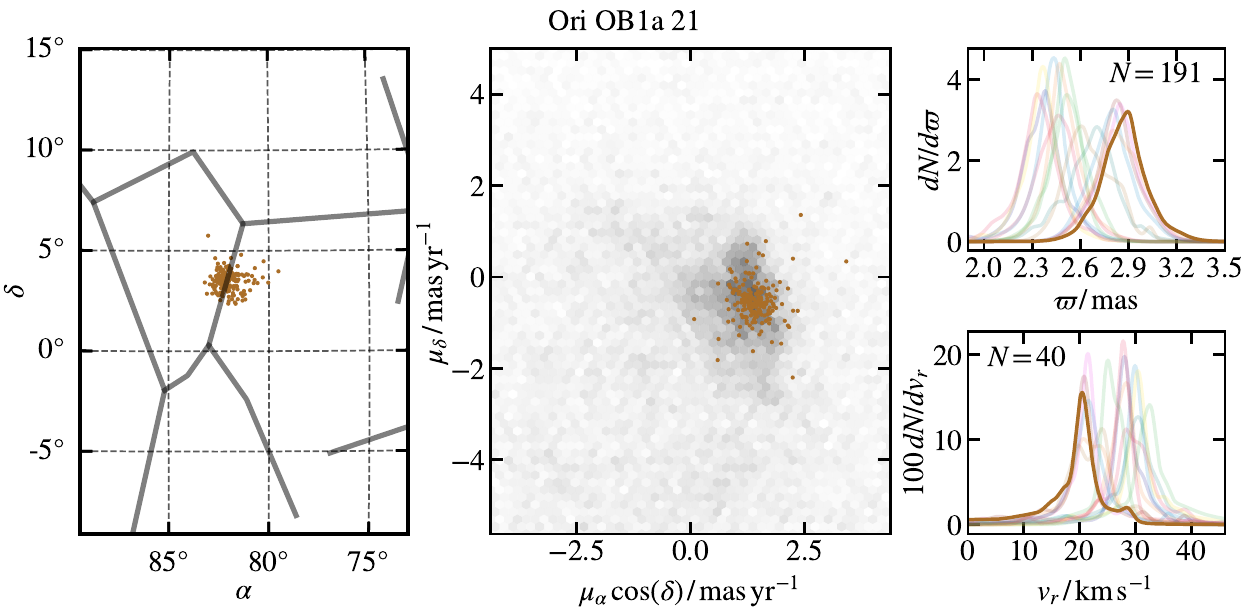}\\
\caption{Positions, proper motions, parallax distribution, and radial velocity distribution for all 15 clusters. }
\label{fig:clusters_a}
\end{figure*}
\addtocounter{figure}{-1}
\begin{figure*}
\centering
\includegraphics[width=0.64\textwidth]{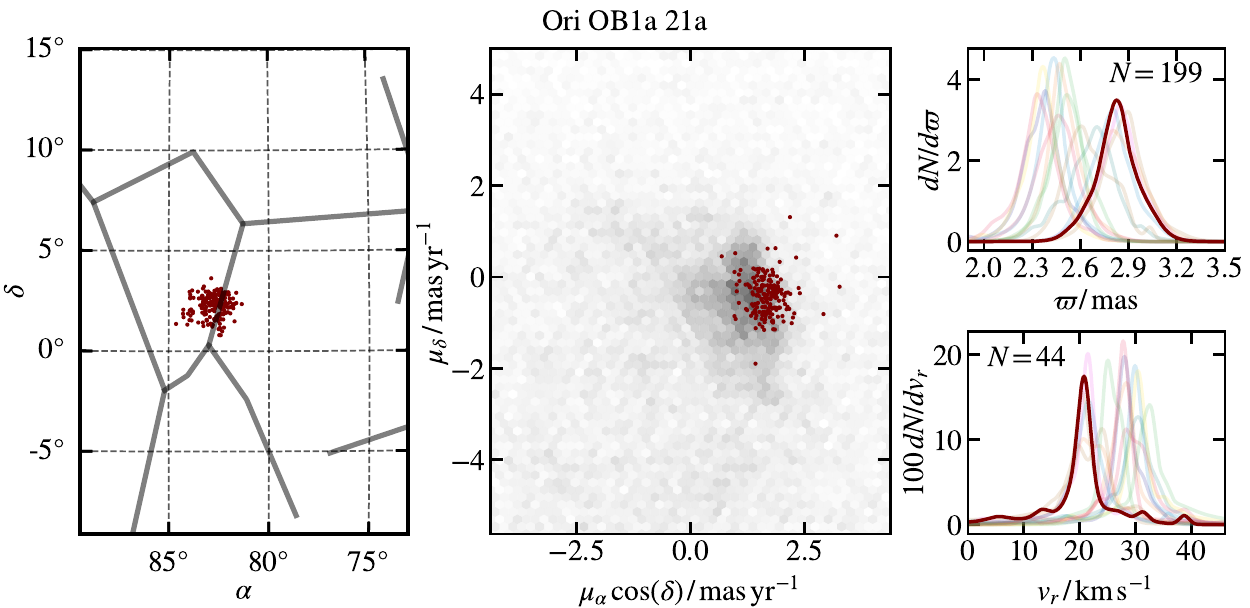}\\[0.3cm]
\includegraphics[width=0.64\textwidth]{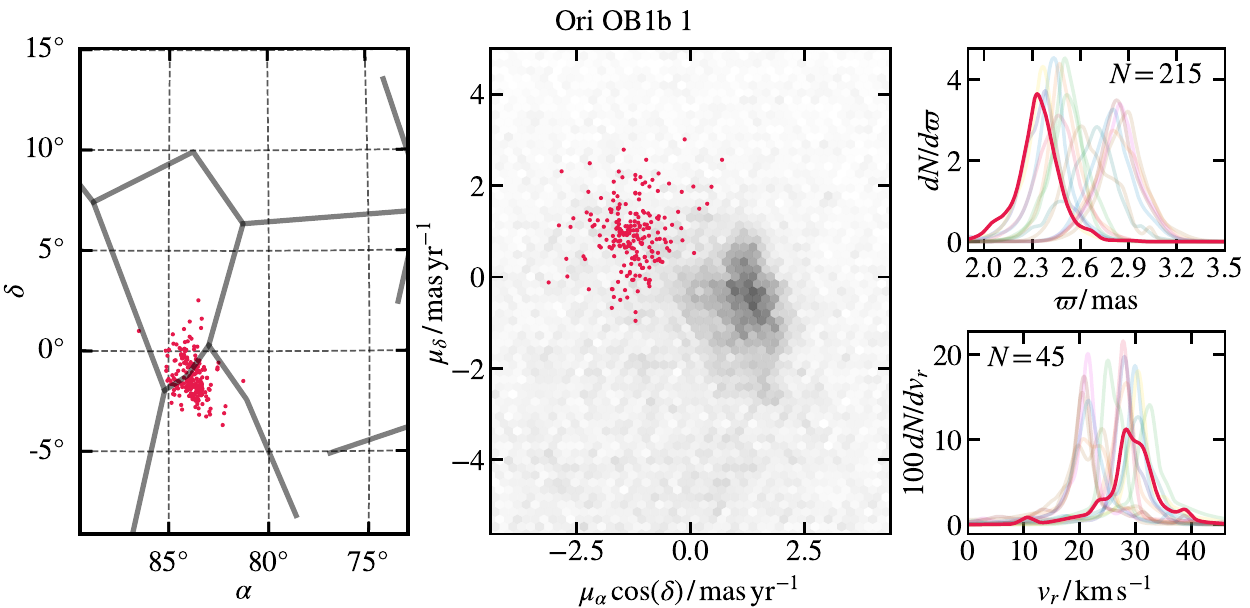}\\[0.3cm]
\includegraphics[width=0.64\textwidth]{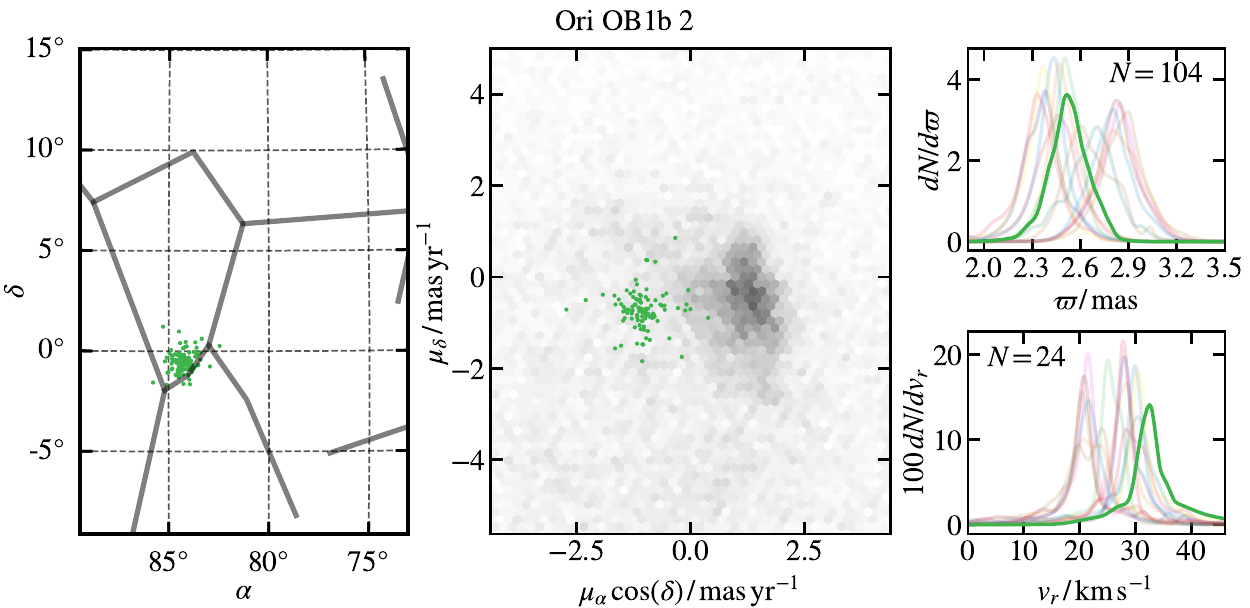}\\[0.3cm]
\includegraphics[width=0.64\textwidth]{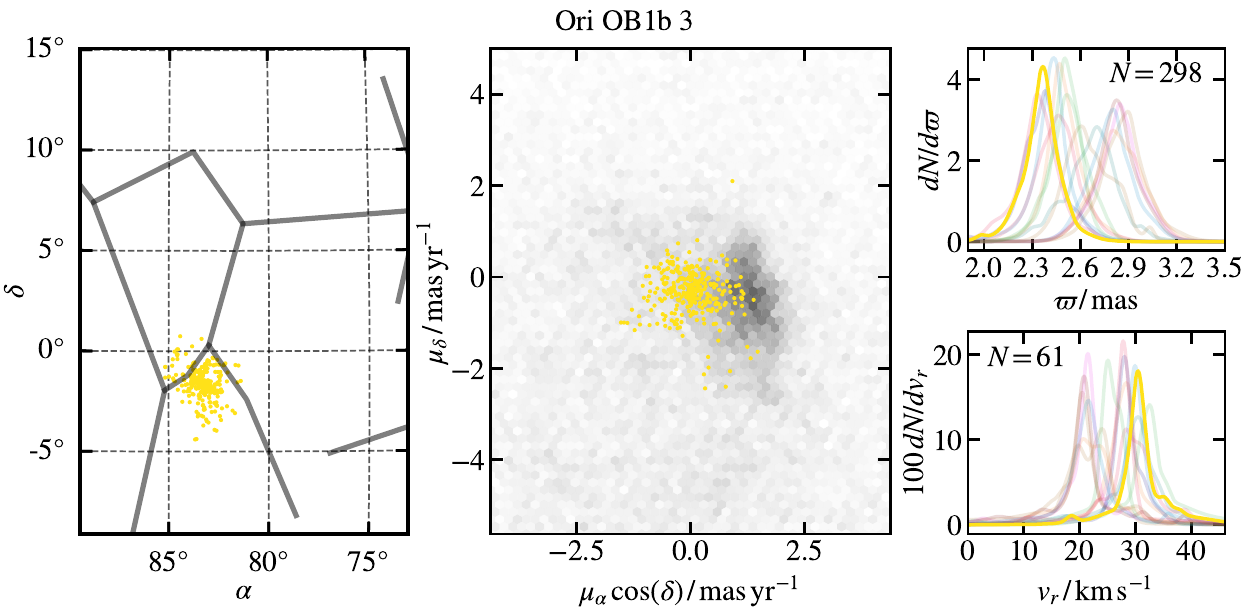}\\
\caption{contd.}
\end{figure*}
\addtocounter{figure}{-1}
\begin{figure*}
\centering
\includegraphics[width=0.64\textwidth]{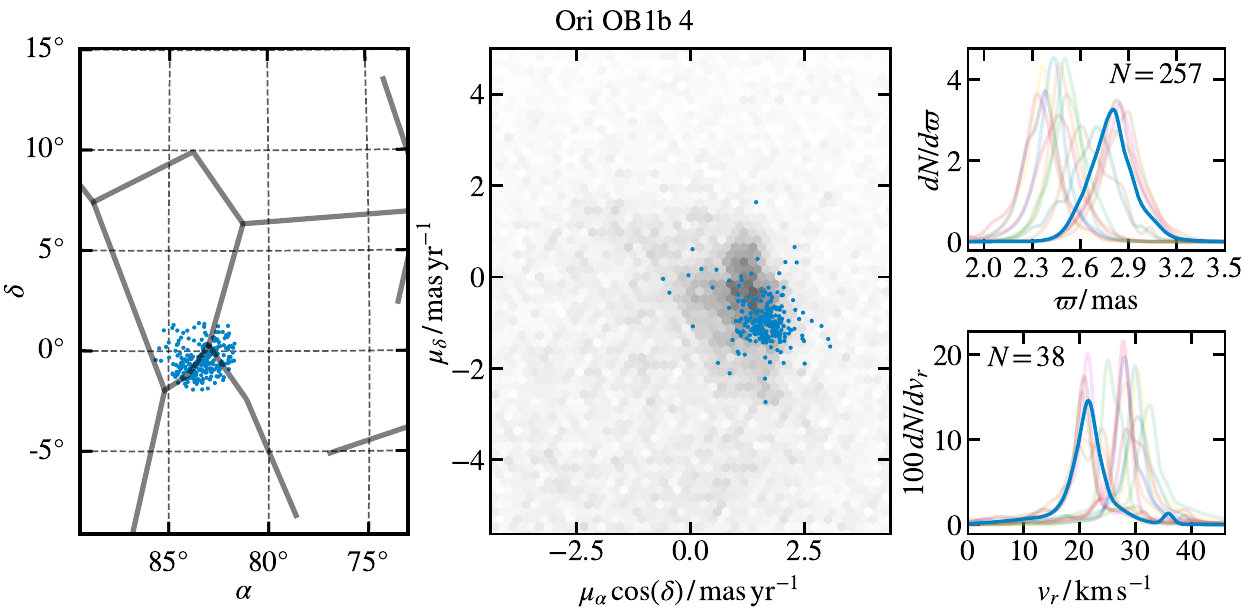}\\[0.3cm]
\includegraphics[width=0.64\textwidth]{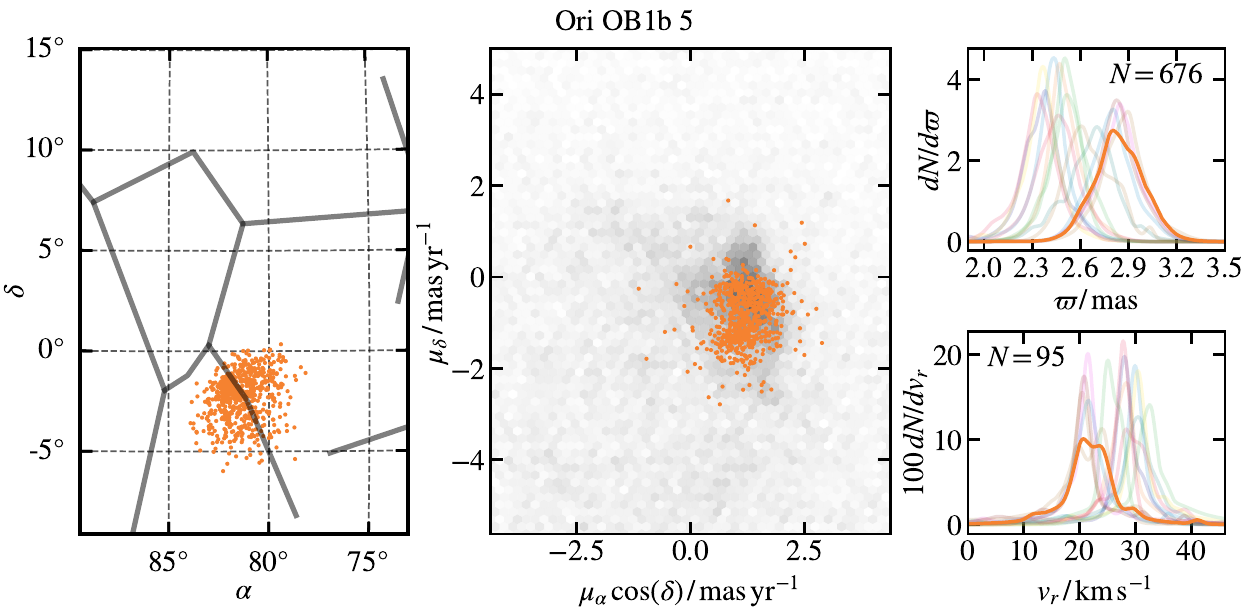}\\[0.3cm]
\includegraphics[width=0.64\textwidth]{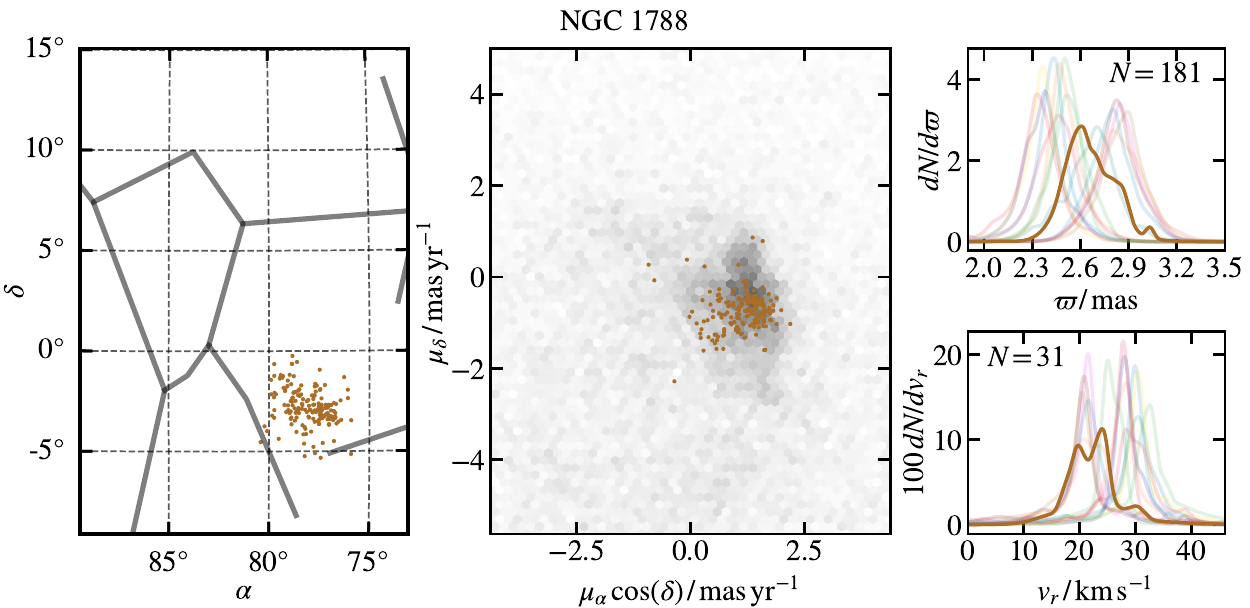}\\[0.3cm]
\includegraphics[width=0.64\textwidth]{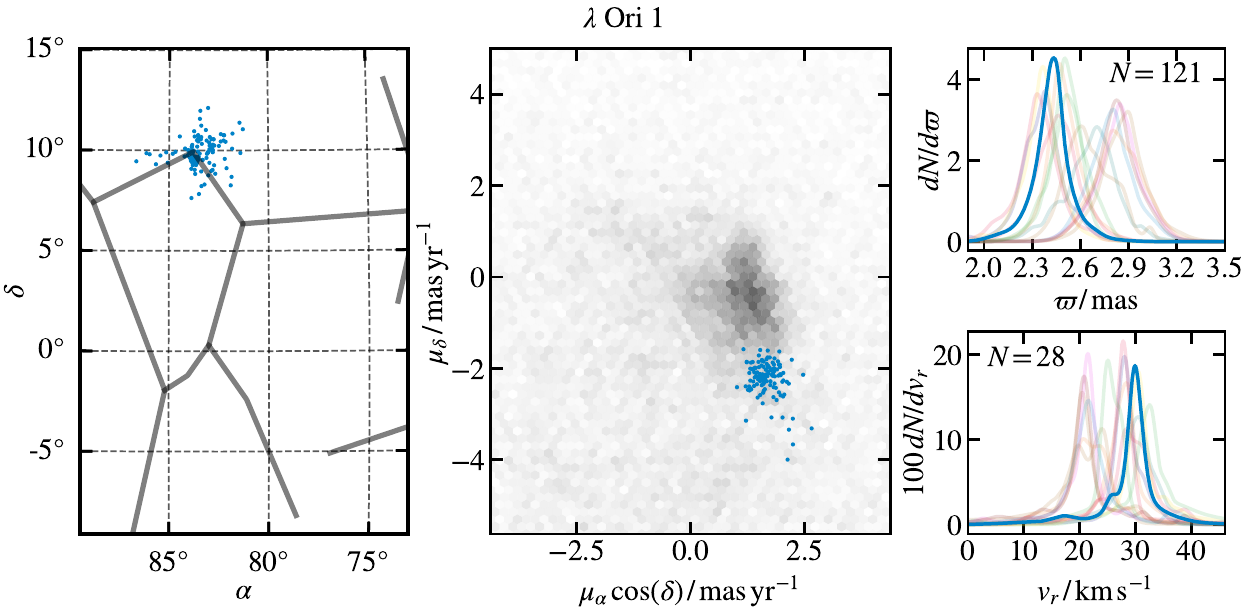}\\
\caption{contd.}
\end{figure*}
\addtocounter{figure}{-1}
\begin{figure*}
\centering
\includegraphics[width=0.64\textwidth]{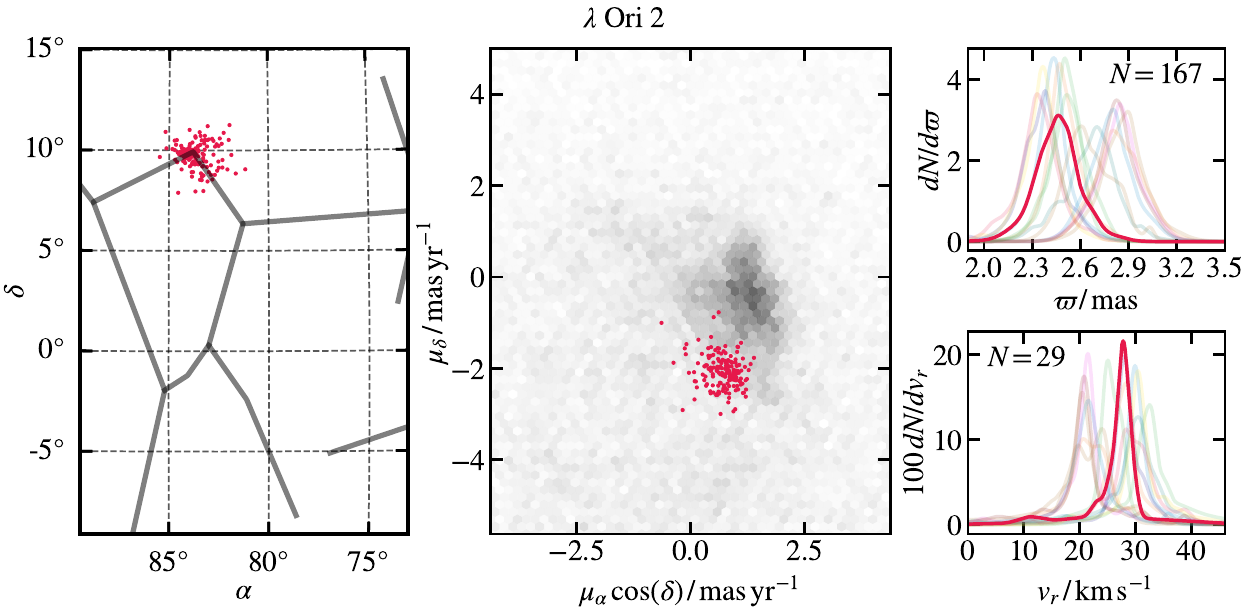}\\[0.3cm]
\includegraphics[width=0.64\textwidth]{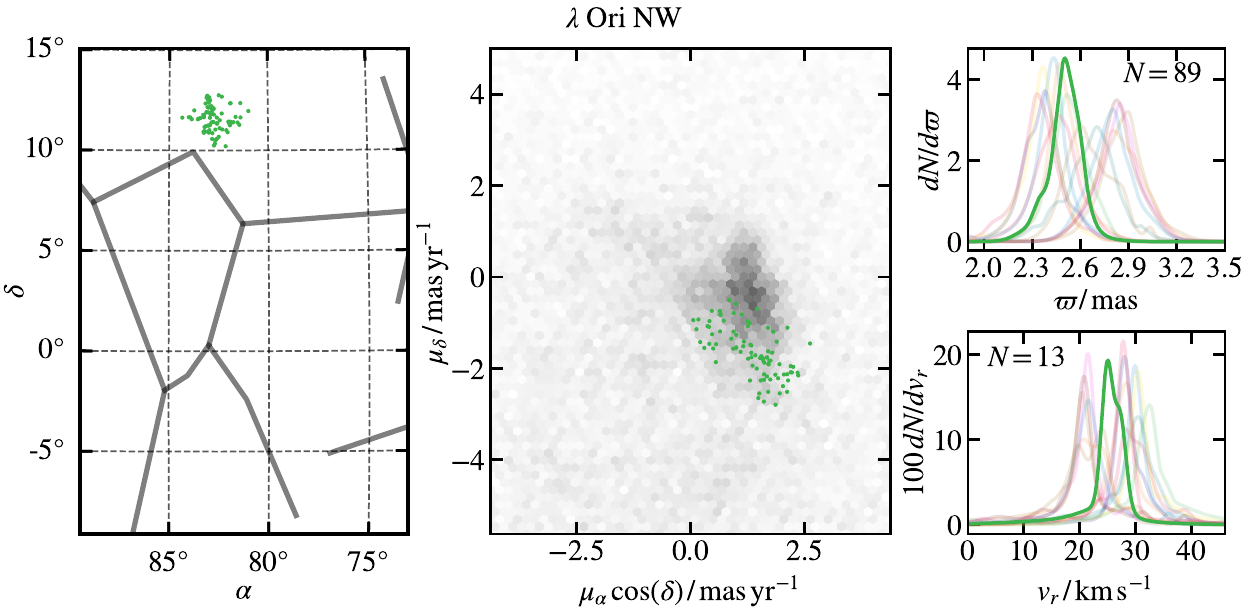}\\[0.3cm]
\includegraphics[width=0.64\textwidth]{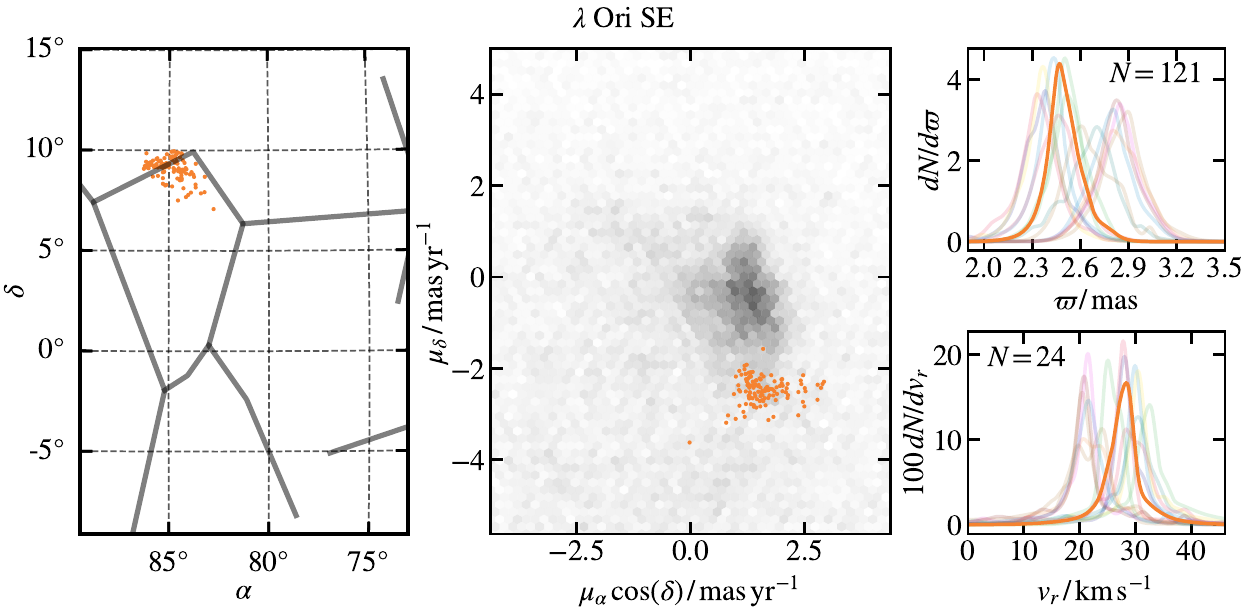}\\
\caption{contd.}
\end{figure*}

\section{HR diagrams}
\label{sec:hr_a}

Figure \ref{sec:hr_a} shows HR diagrams for all 15 clusters made with \textit{Gaia} photometry.

\begin{figure*}
\centering
\includegraphics[width=0.3\textwidth]{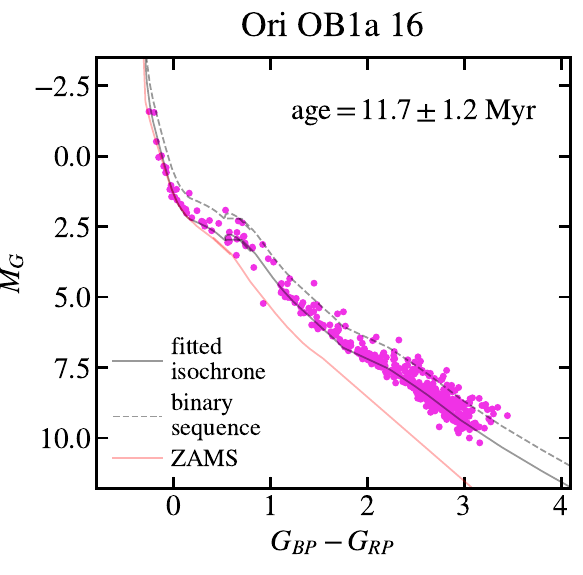}
\includegraphics[width=0.3\textwidth]{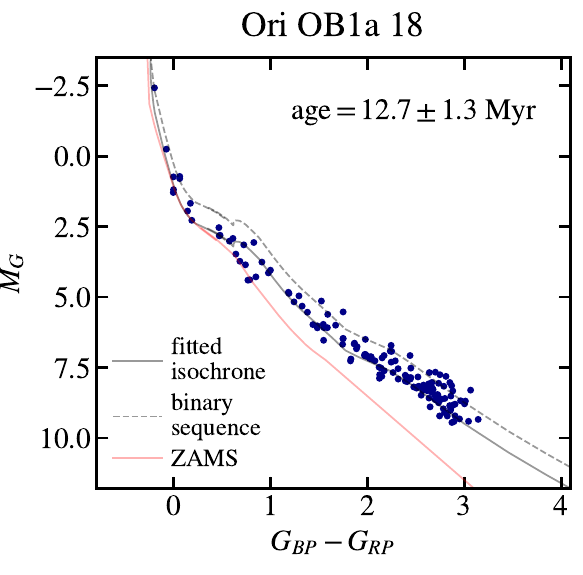}
\includegraphics[width=0.3\textwidth]{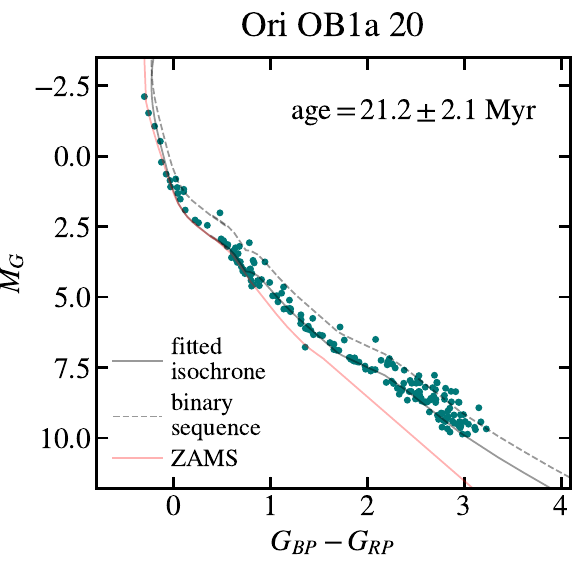}\\[0.3cm]
\includegraphics[width=0.3\textwidth]{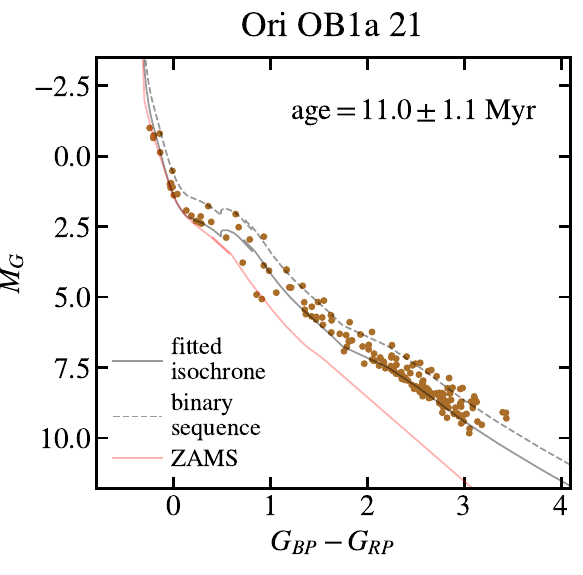}
\includegraphics[width=0.3\textwidth]{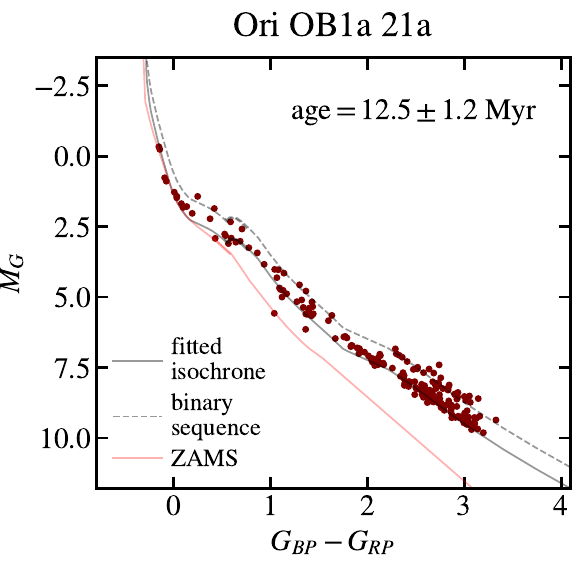}
\includegraphics[width=0.3\textwidth]{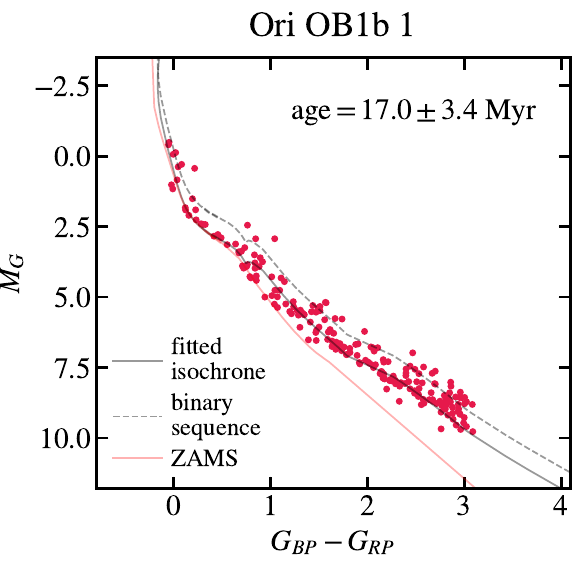}\\[0.3cm]
\includegraphics[width=0.3\textwidth]{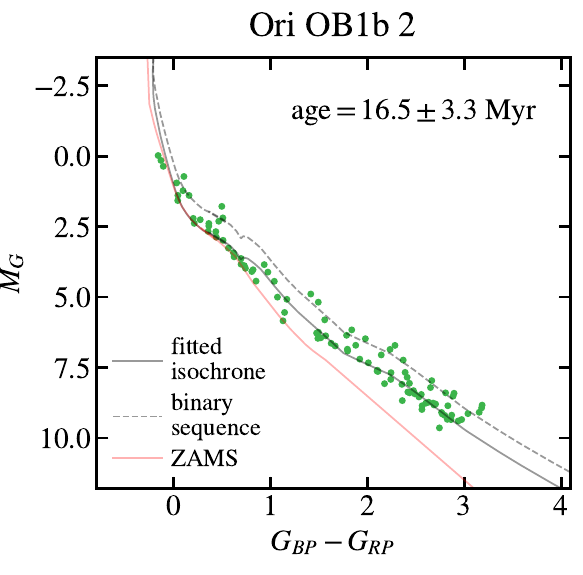}
\includegraphics[width=0.3\textwidth]{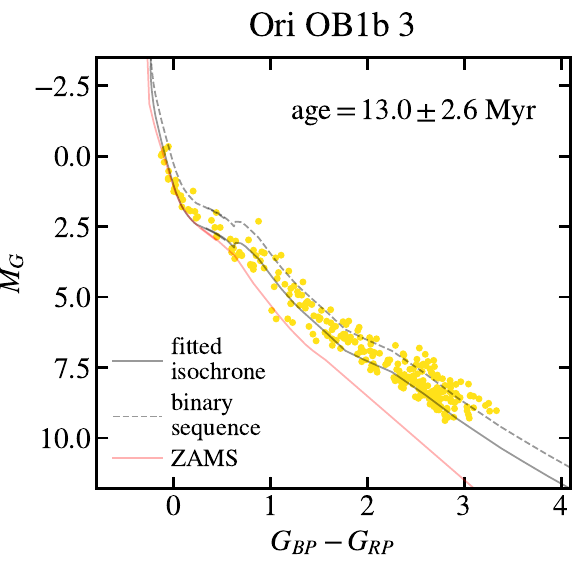}
\includegraphics[width=0.3\textwidth]{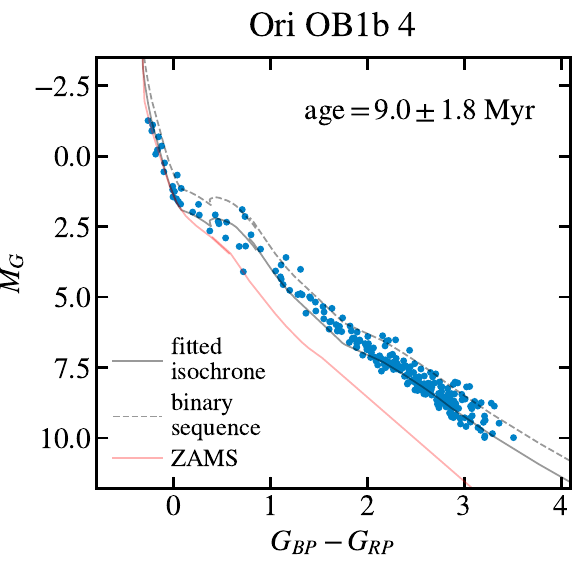}\\[0.3cm]
\includegraphics[width=0.3\textwidth]{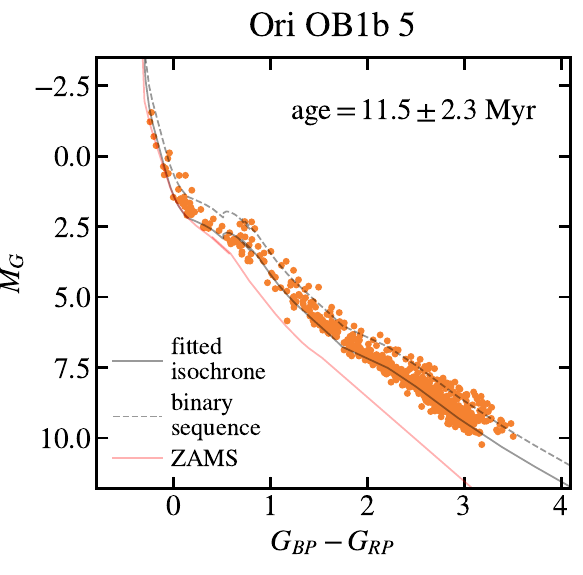}
\includegraphics[width=0.3\textwidth]{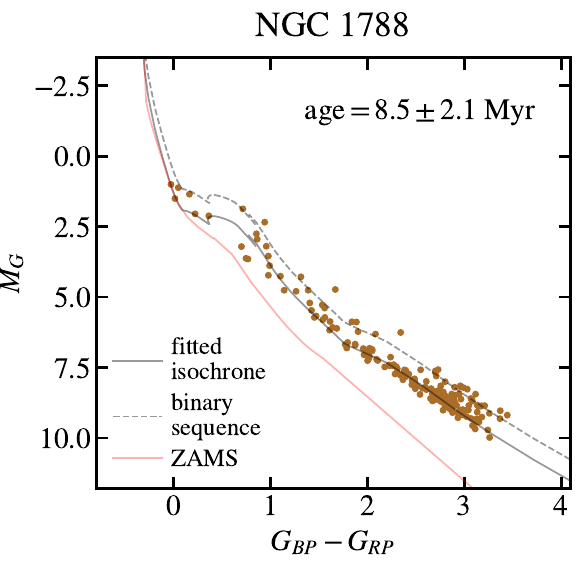}
\includegraphics[width=0.3\textwidth]{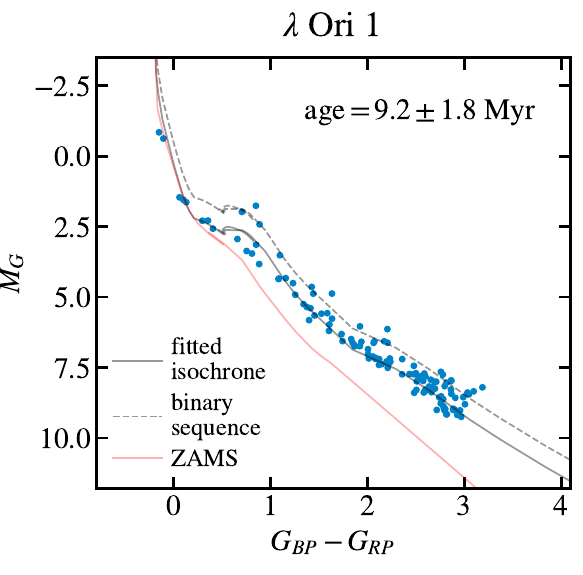}
\caption{HR diagrams of all 15 clusters used in this work. Age as measured by isochrone fitting is given in each panel. Best fitting isochrone is shown as a solid gray line. Dashed line shows its binary sequence. Red line shows the zero-age main sequence (ZAMS).}
\label{fig:hr_a}
\end{figure*}
\addtocounter{figure}{-1}
\begin{figure*}
\centering
\includegraphics[width=0.3\textwidth]{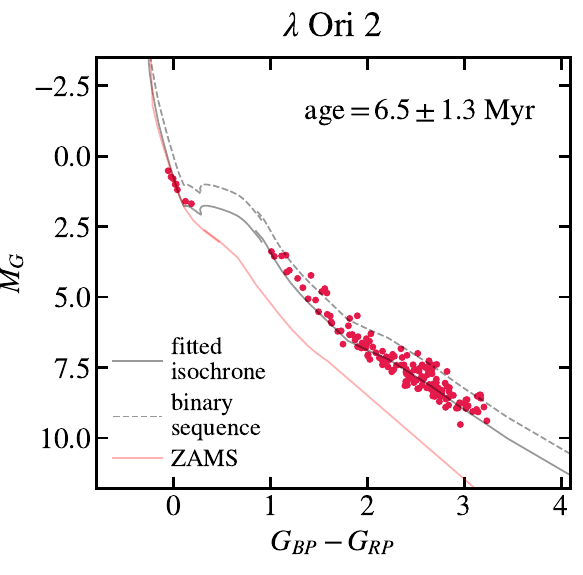}
\includegraphics[width=0.3\textwidth]{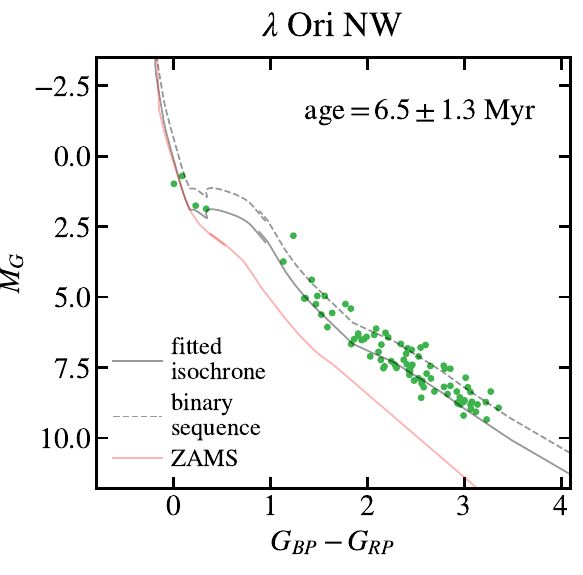}
\includegraphics[width=0.3\textwidth]{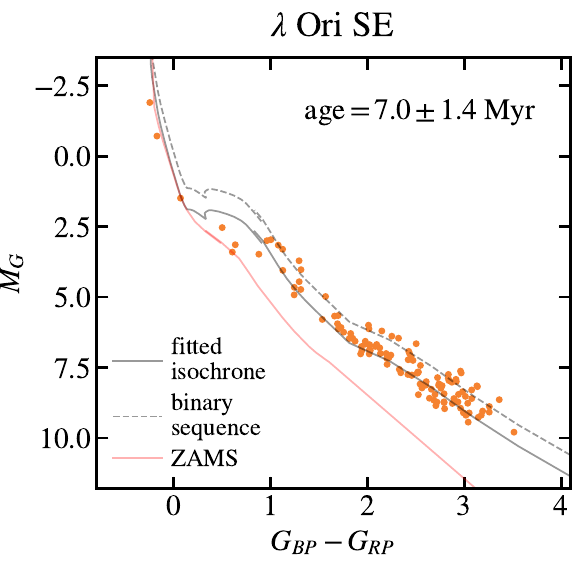}

\caption{contd.}
\label{fig:somelabel}
\end{figure*}

\section{Grid interpolation}
\label{sec:gr_interp}

\begin{figure*}
\centering
\includegraphics[width=0.92\textwidth]{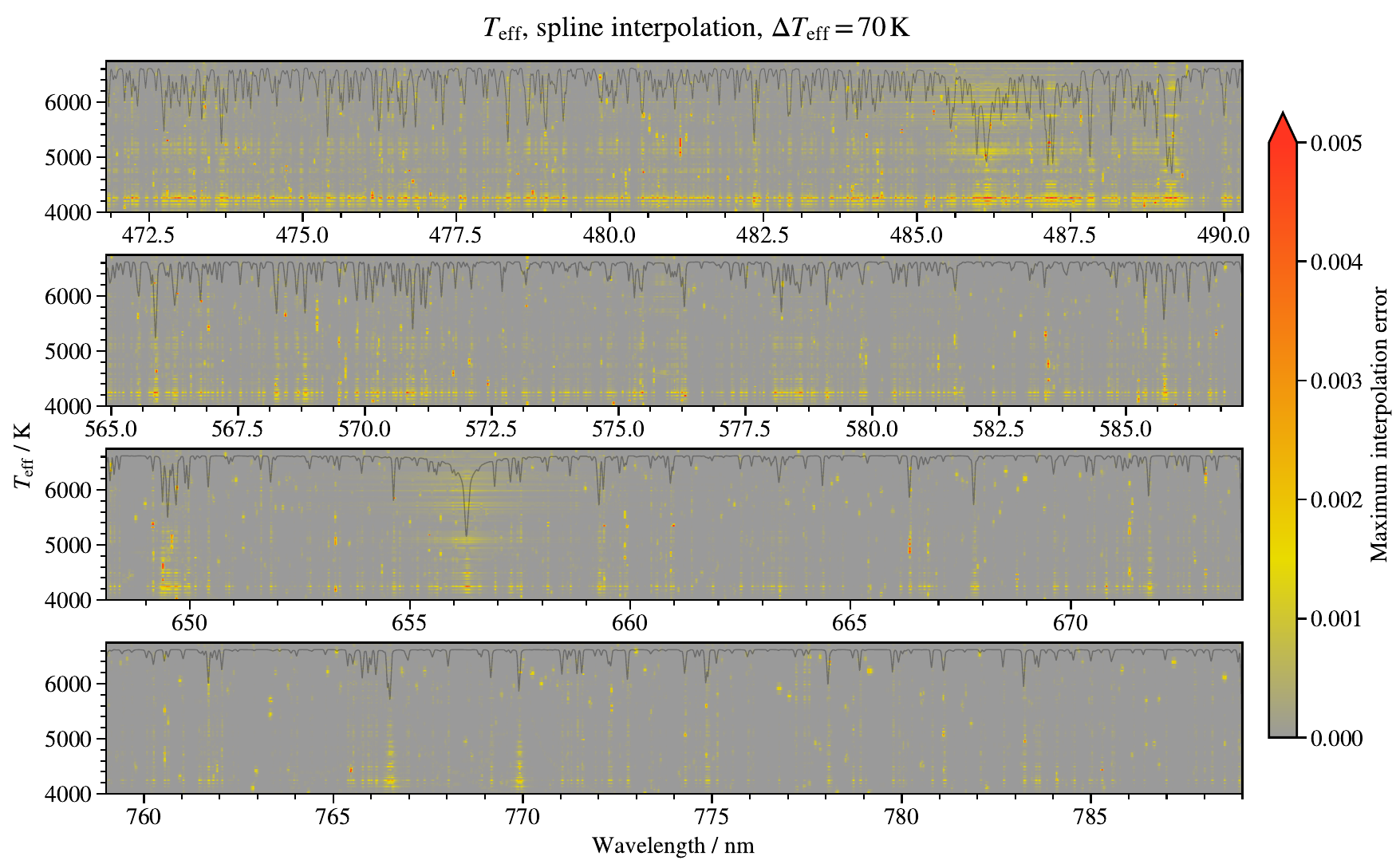}\\
\caption{Maximum interpolation error in the $T_{\mathrm{eff}}$ dimension of the parameter grid. Colours show the difference between a  synthetic spectrum and an interpolated spectrum from a grid. Both spectra are normalized. The interpolated spectrum is always calculated in the middle point between two nodes where the error is expected to be the largest. Errors never exceed the flux of 0.008 (in a normalised spectrum). A spectrum with $T_{\mathrm{eff}}=5250\, \mathrm{K}$, $\log g=4.3$, $[M/H]=-0.05$, $[\alpha/Fe]=0.0$, and $v \sin i=10\,\mathrm{km\, s^{-1}}$ is overploted to illustrate the shape of a typical spectrum.}
\label{fig:inter_teff}
\end{figure*}

\begin{figure*}
\centering
\includegraphics[width=0.92\textwidth]{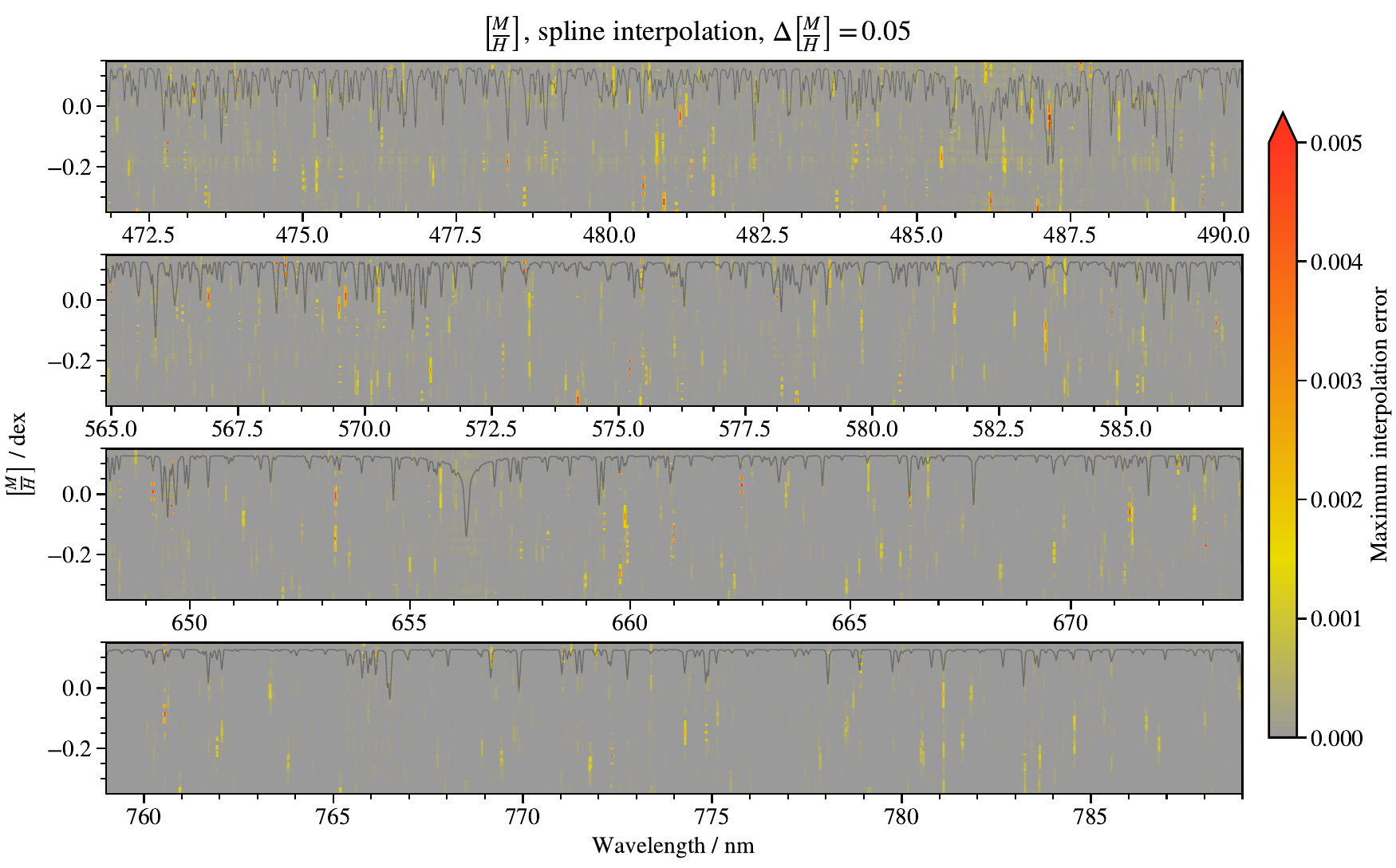}\\
\caption{Same as Figure \ref{fig:inter_teff} but for metallicity.}
\label{fig:inter_mh}
\end{figure*}

\begin{figure*}
\centering
\includegraphics[width=0.92\textwidth]{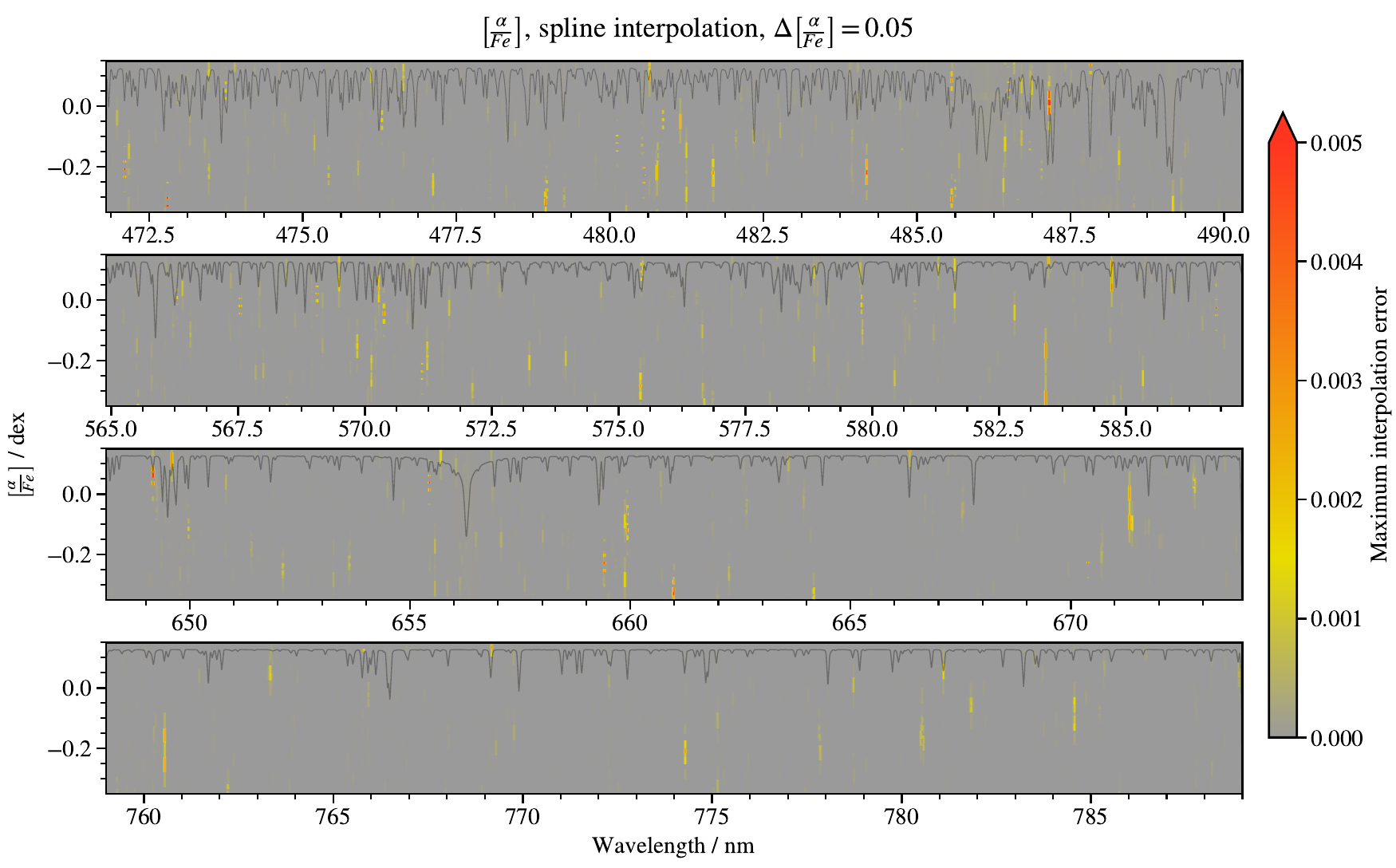}\\
\caption{Same as Figure \ref{fig:inter_teff} but for $\alpha$ abundance.}
\label{fig:inter_alpha}
\end{figure*}

\begin{figure*}
\centering
\includegraphics[width=0.92\textwidth]{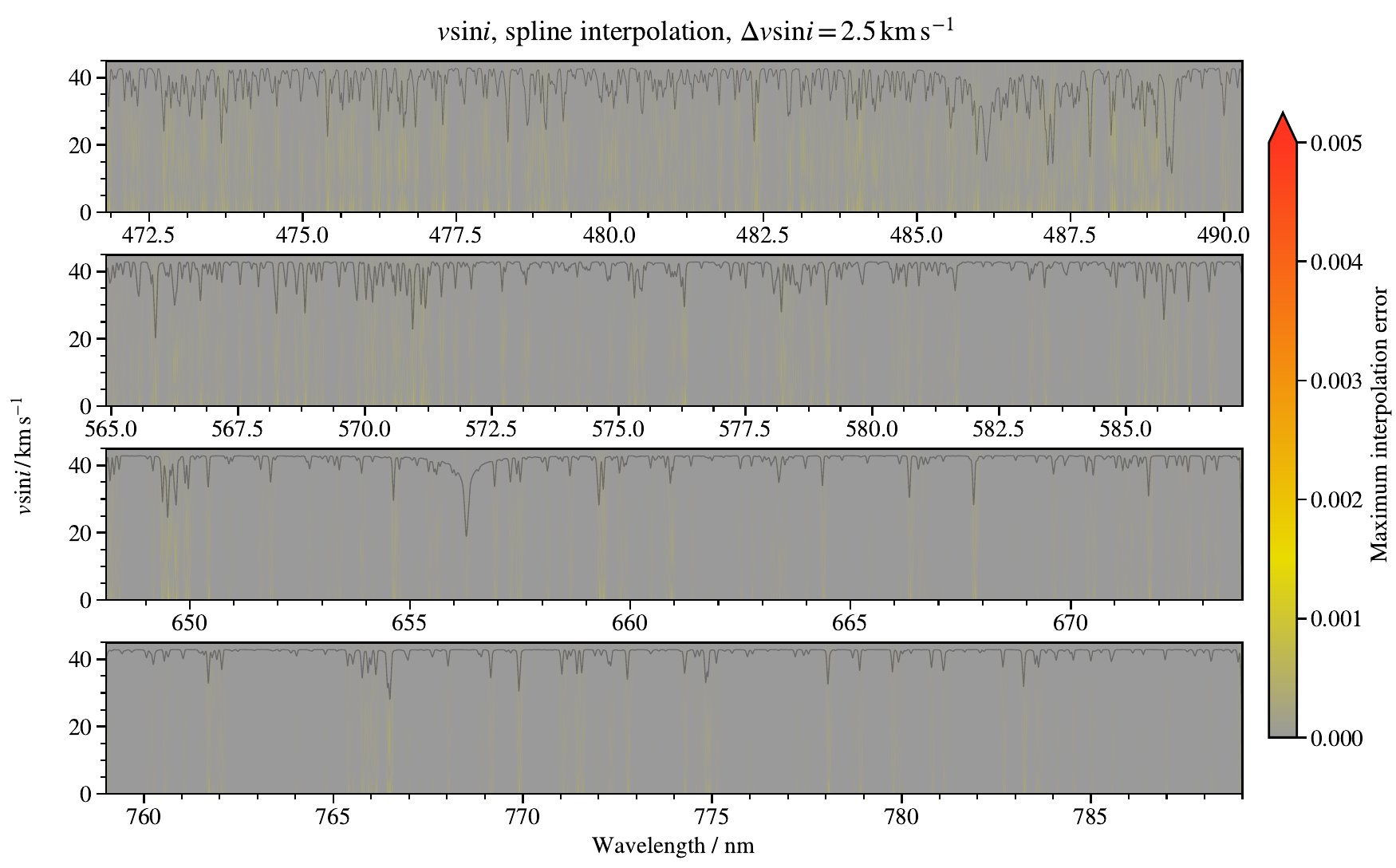}\\
\caption{Same as Figure \ref{fig:inter_teff} but for $v\sin i$.}
\label{fig:inter_vsini}
\end{figure*}

\begin{figure*}
\centering
\includegraphics[width=0.95\textwidth]{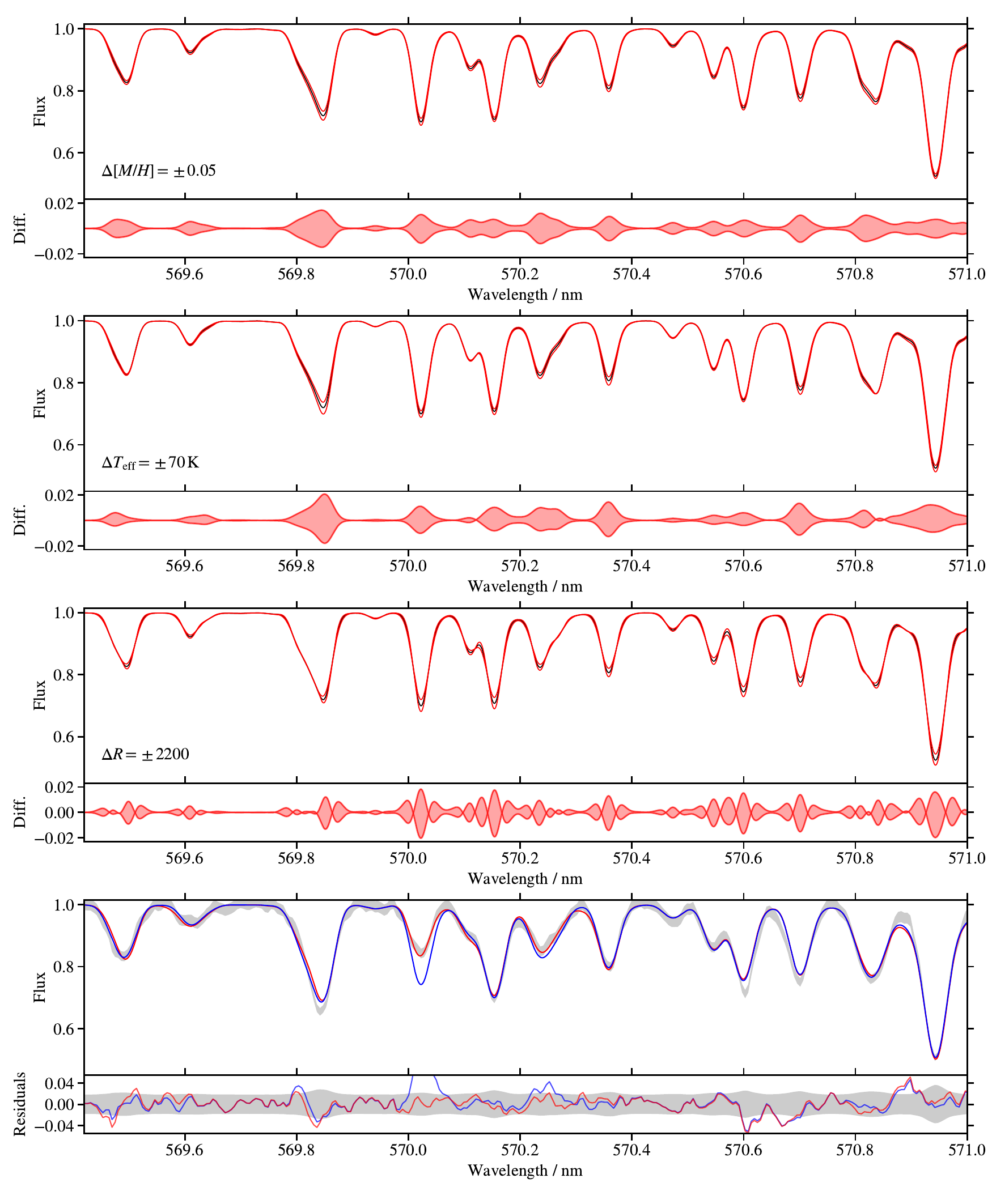}\\
\caption{Sensitivity of synthetic templates to small variations in parameters. First panel: red lines show the difference when metallicity is varied for $\pm 0.05$ dex. Black line shows a spectrum with $T_\mathrm{eff}$=5250 K, $[M/H]$=-0.05, $[\alpha/Fe]$=0.0, $v\, \sin\, i=5\, \mathrm{km\, s^{-1}}$, $R=22\,000$. Adjacent plot shows the differences between red and black lines in more details. Second panel: red lines show the difference when $T_\mathrm{eff}$ is varied for $\pm 70$ K. Black line shows the same spectrum as before. Third panel: red lines show the difference when resolving power $R=\lambda / \Delta \lambda$ is varied for $2200$. Note how a variation in resolution is not degenerated with variations in $[M/H]$ or $T_\mathrm{eff}$. Last panel: Example of an observed spectrum (gray, with thickness corresponding to uncertainty) and best fitting template with a solar mix of elements (blue). Line at 570.02~nm belongs to Cu \textsc{i}, which has much lower abundance than in the Sun. Since the solar mix of elements was used in the fit, the fit deviates significantly for this line. Red line shows the fit with Orion-specific mix of elements (see Table \ref{tab:abund}). }
\label{fig:sens}
\end{figure*}

Figures \ref{fig:inter_teff} to \ref{fig:inter_vsini} demonstrate the interpolation errors introduced by a grid of stellar parameters. Figure \ref{fig:sens} shows a small portion of a spectrum and illustrates the differences between synthetic spectra between grid points. A typical fit of two synthetic spectra to an observed spectrum is also displayed. 

\section{GALAH DR3 abundances}
\label{sec:dr3}

\begin{figure*}
    \centering
    \includegraphics[width=0.95\textwidth]{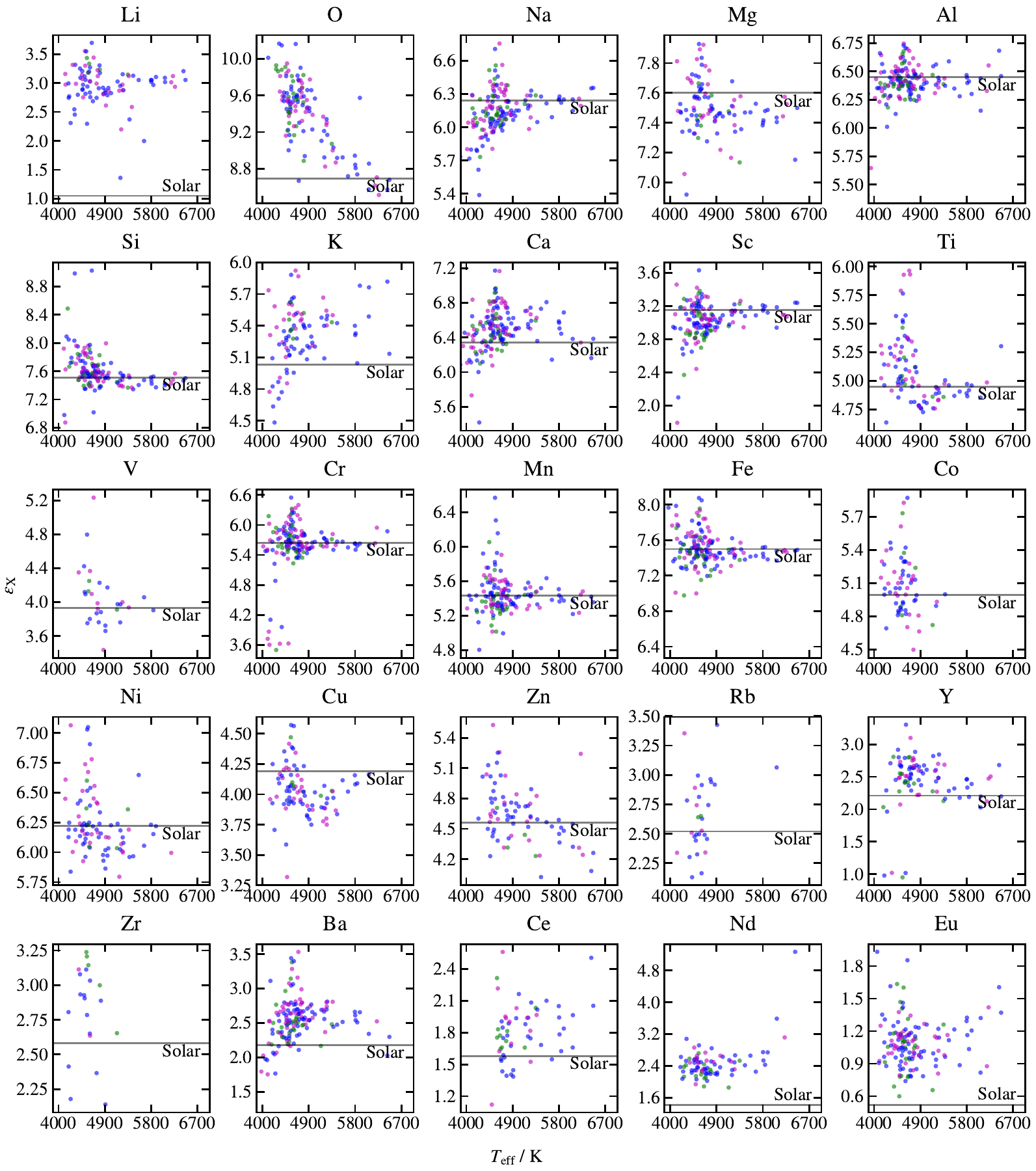}
    \caption{Measured mean abundances of 25 elements given in GALAH DR3 for the same stars as in Figure \ref{fig:detrend_elements}. Stars without measured abundances in GALAH DR3 are missing from this plot.}
    \label{fig:detrend_elements_dr3}
\end{figure*}

Figure \ref{fig:detrend_elements_dr3} shows mean abundances of 25 elements for a cross-match between our study and GALAH DR3.

\bsp	
\label{lastpage}
\end{document}